\documentclass[conference]{IEEEtran}
\IEEEoverridecommandlockouts
\usepackage{cite}
\usepackage{amsmath,amssymb,amsfonts}
\usepackage{graphicx}
\usepackage{textcomp}
\usepackage{xcolor}
\usepackage [english]{babel}
\usepackage [autostyle, english = american]{csquotes}
\MakeOuterQuote{"}
\usepackage{enumitem}
\usepackage{graphicx}
\usepackage{float}
\usepackage{dsfont}
\usepackage{graphics} % for pdf, bitmapped graphics files
\usepackage{epsfig} 
\usepackage{tabularx}
\usepackage{multicol}
\usepackage{multirow}
\usepackage{tabularx}
\usepackage{pdflscape}
\usepackage{bbm}
\usepackage[noindentafter]{titlesec}
\usepackage{bbding}
\usepackage{subfigure}
\usepackage{wasysym}
\usepackage{amssymb}% http://ctan.org/pkg/amssymb
\usepackage{pifont}% http://ctan.org/pkg/pifont
\usepackage{amssymb}
\usepackage{afterpage}
\usepackage{capt-of}
\usepackage{bbm}
\usepackage{rotating}
\usepackage{lscape}
\usepackage{booktabs}
\usepackage{hyperref}
\usepackage{pgfplots}
\pgfplotsset{width=8cm,compat=1.9}
\graphicspath{ {images/} }
\usepackage{amsmath}
\usepackage{braket} % needed for \Set
\usepackage{algorithm}
\usepackage{xcolor}
\usepackage{mathtools}
\DeclarePairedDelimiter\norm{\lVert}{\rVert}
\usepackage[noend]{algpseudocode}
\usepackage{mathtools,bm}
\newcommand{\blue}[1]{\textcolor{black}{#1}}

\def\BibTeX{{\rm B\kern-.05em{\sc i\kern-.025em b}\kern-.08em
    T\kern-.1667em\lower.7ex\hbox{E}\kern-.125emX}}

\DeclareMathOperator*{\argmin}{arg\,min}    
    
\makeatletter
\patchcmd{\@maketitle}
  {\addvspace{0.5\baselineskip}\egroup}
  {\addvspace{-1\baselineskip}\egroup}
  {}
  {}
\makeatother

\begin{document}
\bstctlcite{IEEEexample:BSTcontrol}

\title{DeepFT: Fault-Tolerant Edge Computing using a Self-Supervised Deep Surrogate Model}

\author{
        % \IEEEauthorblockN{Anonymous Authors}%
  \IEEEauthorblockN{Shreshth Tuli\IEEEauthorrefmark{1}, Giuliano Casale\IEEEauthorrefmark{1}, Ludmila Cherkasova\IEEEauthorrefmark{2} and Nicholas R. Jennings\IEEEauthorrefmark{3}}
 \IEEEauthorblockA{{\IEEEauthorrefmark{1}Department of Computing, Imperial College London, UK. Email: \{s.tuli20,g.casale\}@imperial.ac.uk}.}
  \IEEEauthorblockA{{\IEEEauthorrefmark{2}ARM Research, USA. Email: ludmila.cherkasova@arm.com}.}
  \IEEEauthorblockA{{\IEEEauthorrefmark{3}Loughborough University, UK. Email: n.r.jennings@lboro.ac.uk}.}
}

\maketitle
\thispagestyle{plain}
\pagestyle{plain}

% Data rising in edge and pervasive. Automated labelling not accurate, expert labelling not feasible. Prior approaches fail in volatile settings. Need adaptive and proactive approach. DeepFT develops upon a deep surrogate model with prototypical networks and transformers to proactively detect faults. DeepFT leverages self-supervised learning setup using co-simulations, and optimization to input to optimize QoS. 

\begin{abstract}
%The emergence of latency-critical AI applications has given rise to a frequent adoption of large-scale paradigms like pervasive edge computing. 
The emergence of latency-critical AI applications has been supported by the evolution of the edge computing paradigm. However, edge solutions are typically resource-constrained, posing reliability challenges due to heightened contention for compute and communication capacities and faulty application behavior in the presence of overload conditions. Although a large amount of generated log data can be mined for fault prediction, labeling this data for training is a manual process and thus a limiting factor for automation. Due to this, many companies resort to unsupervised fault-tolerance models. Yet, failure models of this kind can incur a loss of accuracy when they need to adapt to non-stationary workloads and diverse host characteristics. To cope with this, we propose a novel modeling approach, called DeepFT, to proactively avoid system overloads and their adverse effects by optimizing the task scheduling and migration decisions. DeepFT uses a deep surrogate model to accurately predict and diagnose faults in the system and co-simulation based self-supervised learning to dynamically adapt the model in volatile settings. It offers a highly scalable solution as the model size scales by only 3 and 1 percent per unit increase in the number of active tasks and hosts.
% Moreover, DeepFT uses co-simulations to adapt the surrogate model in a self-supervised fashion and uses backpropagation to input to optimize the scheduling decisions. 
Extensive experimentation on a Raspberry-Pi based edge cluster with DeFog benchmarks shows that DeepFT can outperform state-of-the-art baseline methods in fault-detection and QoS metrics. Specifically, DeepFT gives the highest F1 scores for fault-detection, reducing service deadline violations by up to 37\% while also improving response time by up to 9\%.
\end{abstract}

\begin{IEEEkeywords}
Self-supervised learning, Task Migration, Surrogate Models, Fault-Tolerance, Edge Computing.
\end{IEEEkeywords}
%%%%%%%%%%%%%%%%%%%%%%%%%%%%%%%%%%%%%%%%%%%%%%%%%%%%%%%%%%%%%%%%%%%%%%%%%%%%%%%%
\section{Introduction}\label{sec:introduction}
\noindent
% What is pervasive edge computing, why it is popular. Modern applications changing to AI. Big data platforms with large-scale IoT devices generating enormous amount of data. Edge devices resource constrained, tend to get overloaded. Need for fault tolerance.
Fault-tolerant computing is an essential aspect of reliable service delivery. The recent IoT device explosion has given rise to a significant increase in sensing data. In such settings, sending all data to cloud backends is infeasible, compelling us to process data at the edge~\cite{narayanan2020key}. \blue{Resource constraints in edge devices can lead to unreliability, subsequently} causing information loss and poor system performance. %Unreliability in such pervasive architectures rises from the transition of applications to being more compute intensive as well as the severe resource constraints in edge devices~\cite{cox2021masa}. Resource-hungry applications often lead to resource contention and system overload, giving rise to performance degradation and application failure~\cite{vasilakos2020towards}. 
For example, the processing nature of data-intensive applications and bursty request arrivals often lead to resource contention and system overload, causing application performance degradation and failures~\cite{vasilakos2020towards,tuli2022dragon}. 
This is exacerbated by the tight Quality of Service (QoS) requirements of modern applications, for instance, tight execution budgets and time-sensitive tasks. To alleviate the adverse effects of system overload on QoS, fault-tolerant schemes are required for reliable service delivery.

% Such data-centric architectures of pervasive computing rely on spontaneous data sharing and processing in the peer network to improve service latency and reduce costs.

% lack of data, volatile workloads, 
\textbf{Challenges.} The key challenge tackled in this paper is to devise a proactive overload and contention protection scheme that is efficient and accurate in the face of a lack of labeled training data. As most traditional schemes depend on supervised training, most edge computing environments lack fault labels. Recovery steps are required that deal with the diverse effects of resource contention, such as network packet drops, memory errors or disk failures requiring different remediation steps. Besides, for such recovery steps to be effective, faults must be predicted {\it beforehand} as reactive schemes have been shown to have lower efficacy in edge systems~\cite{ray2020proactive}. We need  proactive predictions to enable running remediation steps in advance to conform with the near real-time requirement of fault recovery in volatile environments. % with non-stationary workloads demands, host resource capacities or user service requirements.  
In such environments, the statistical moments and correlations of the workload characteristics are non-stationary and vary over time, requiring continuous task re-scheduling. This further complicates the distinction between workload dynamism and fault-based deviation in measurements. %~\cite{ristov2020resilient}. %Traditional methods mainly rely on manual labeling of faults through domain experts, which is infeasible in modern IoT solutions with enormous amounts of log data~\cite{awgg}. 
As the number of compute tasks and available processing devices increases, this also becomes more challenging to solve in a scalable manner. Further, to avoid the overheads resulting from incorrect remediation decisions, the designed proactive schemes have to be parsimonious in executing such actions. This means, to tackle challenges such as the lack of data and environment volatility, we need an accurate unsupervised method to proactively predict and remediate the faults and adapt to changing settings. 

% existing supervised methods need labelled data, existing online unsupervised reconstruction based, get fault score after execution (reactive not proactive). 
\textbf{Existing solutions.} Over the past few years, several fault-tolerance approaches have been proposed that use heuristics or traditional supervised learning strategies~\cite{samanta2021fault, cmodlb, dftm}. Many such approaches utilize redundant nodes as a fallback to deal with faults. However, in edge computing environments, having node redundancy is inefficient and counterproductive for low-cost deployments~\cite{bagchi2019dependability, goudarzi2020application}. Other prior work mainly aims to recover from a faulty state that occurs due to resource contention and over-utilization. This is common in scenarios where resource-intensive applications are deployed on constrained edge computing environments. Due to the limited capacities of edge nodes, most works aim at solving the fault-tolerance problem by balancing the load across different nodes in the system~\cite{kumar2015fault} using preemptive migrations. Such migration decisions checkpoint the running instances of tasks, migrate the saved states to other nodes, and restore their execution states. As this can be performed for a subset of tasks in the system in a non-blocking fashion, it is an efficient method for load balancing in edge or cloud platforms. However, choosing the tasks to migrate and the destination host nodes for these tasks is challenging~\cite{ledmi2018fault}. Several classical methods use threshold-based heuristics to select faulty hosts, then determine the most resource-hungry tasks within such hosts to be migrated to other nodes~\cite{beloglazov2012optimal}. Other previously proposed methods use a supervised learning framework with  AI models to achieve the same~\cite{eclb, cmodlb, tuli2021pregan}. However, the supervised learning schemes are not directly suitable for large-scale settings due to the pragmatic difficulties in obtaining labeled data~\cite{awgg}. Hence, recent methods offer unsupervised schemes like  clustering~\cite{pcft}, sparse neural networks~\cite{awgg}, autoencoders~\cite{topomad} or other meta-heuristic formulations~\cite{pcft, dftm}. \blue{Such methods extract patterns from historical log data without any fault labels, mostly using data reconstruction techniques.} However, such methods struggle to adapt in volatile environments as they need several datapoints corresponding to operational and performance characteristics of the system~\cite{topomad}. Moreover, these methods wait for the latest data to detect faults, diminishing their appeal. %Thus, we propose a novel fault-tolerance model and demonstrate its efficacy against the state-of-the-art deep learning based baselines~\cite{awgg, topomad}. 

% unsupervised and self-supervised, bpti, co-simulations, multi head attn
\textbf{Background and new insights.} \blue{A drawback of not using any human domain knowledge is that unsupervised learning approaches can vastly deviate from the ground truth~\cite[\S~6.5]{pecht2019machine}. A class of unsupervised learning, called self-supervised learning, is a refactored supervised scheme where the labels are generated by the model itself~\cite{goodfellow2016deep}. % This approach relies on robust supervised methods with self-generated data labels.
However, without a trained model, generating such labels is hard. Thus, we propose to obviate the lack of supervised data by means of online simulation, enabled and driven by system data (as per a digital modeling twin~\cite{santos2020use,tuli2022simtune}).  A co-simulated digital twin, referred to as a co-simulator in the rest of the discussion, is a software that models the behavior of a physical system, which in our case is an edge computing platform. Recent developments, in co-simulator design, enable a model to quickly run discrete-event simulations to get an estimate of a future state of the system, allowing proactive fault remediation~\cite{tuli2021cosco}. However,  running a co-simulator for every preemptive migration decision is computationally expensive. A key ingredient in our work is the use of a surrogate model that mimics the behavior of these expensive simulations~\cite{kochenderfer2019algorithms}. Unlike a co-simulator, our surrogate model captures the fundamental aspects of the system dynamics observed from the data without incurring a state space explosion.} We propose to use a co-simulator to generate self-supervision labels to fine-tune a reconstruction-based model on-the-fly for fault detection. Using self-generated data permits model training with limited historical data and adaptation to dynamic settings. To obviate volatility, we cast the problem as a few-shot learning classification~\cite{wang2020generalizing}, whereby a small number of samples are needed to train a neural network. This allows for faster learning under highly volatile workloads. %Each of these different insights cannot be directly used without intelligently integrating each component with necessary adaptations, which is done in this work. 

\textbf{Our contributions.} In this work, we propose \underline{DeepFT}\footnote{We shall release the code on GitHub upon acceptance.}, a  scalable \underline{Deep} surrogate model based \underline{F}ault-\underline{T}olerance method. DeepFT utilizes a neural network as a surrogate model to predict a future system state, the possible presence of faults affecting it and their class. It uses co-simulation to generate fault labels and trains the surrogate in a few-shot setting. With a trained surrogate model, DeepFT generates preemptive migration decisions for fault tolerance. Extensive experiments on a Raspberry-Pi based edge cluster show that DeepFT \blue{outperforms five recently proposed different baselines} in terms of fault-detection and QoS metrics. Specifically, DeepFT reduces response times and latency-based service level objective (SLO) violations by 9.41\% and 37.21\%, respectively, thanks to its higher fault prediction scores and lower overheads.

% \textbf{Our contributions.} In this work, we propose \underline{DeepFT}, a \underline{Deep} surrogate model based \underline{F}ault-\underline{T}olerance method. DeepFT uses a neural network as a surrogate model to predict a future system state, the possible presence of faults affecting it and their class. Using this, it generates preemptive migration decisions. Recent advances of attention operations, allow DeepFT to scale with the number of hosts or active tasks in the system.  
% DeepFT is the \textit{first} method that uses co-simulated next state prediction with reconstruction based fault score evaluation and few-shot classification to proactively predict migration decisions. This allows DeepFT to avoid faults without explicitly labelling faulty system states. DeepFT is trained using a pre-collected trace data and fine-tuned using a co-simulator to adapt and generalize to unseen settings.  Using reconstruction error as a fault likelihood score, DeepFT uses the surrogate model to optimize the scheduling decision using gradients with respect to input~\cite{tuli2021cosco}. Gradient-based optimization is shown to be more time-efficient and converge to improved objective values. Extensive experiments on a Raspberry-Pi based edge cluster show that DeepFT performs \textit{best} in terms of fault-detection and QoS metrics. Specifically, DeepFT reduces response times and latency based service level objective (SLO) deadline violations by 9.41\% and 37.21\%, respectively, thanks to its higher fault prediction scores and lower overheads.

The rest of the paper is organized as follows. Section~\ref{sec:related_work} overviews related work. Section~\ref{sec:method} outlines the DeepFT methodology for model training and the optimization of the scheduling decisions. A performance evaluation of the proposed method is presented in Section~\ref{sec:experiments}. Finally, Section~\ref{sec:conclusion} concludes the paper and enumerates future directions.

\section{Related Work}
\label{sec:related_work}
\noindent

We now analyze the prior work in fault-tolerant computing. 

\textbf{Meta-Heuristic Methods.} Most contemporary fault prevention techniques employ some form of heuristics or machine learning models. Dynamic Fault-Tolerant Migration (DFTM)~\cite{dftm} proposes a recovery mechanism through live virtual machine (VM) migration in cloud computing environments. This work presents an Integer Linear Programming (ILP) formulation to generate an optimal live-migration decision.  However, the ILP model is not scalable due to its state explosion, limiting its efficacy in large-scale edge deployments. Another related work for cloud computing environments is the Proactive Coordinated Fault Tolerance (PCFT)~\cite{pcft} method, which uses Particle Swarm Optimization (PSO) to solve the coordinated consensus issue in fault-tolerance with multiple heterogeneous cloud nodes. The consensus reached at the end of this method is a VM migration decision to optimize network consumption, transmission overheads and the service response times. Like other heuristic-based methods (e.g., MCAFM and PAFM~\cite{ray2020proactive}), this method uses a predefined thermal map of the hosts' CPU to determine source hosts and PSO based QoS optimization to decide target hosts. Then, PCFT migrates the VM with the maximum transmission overhead to the target host. However, PCFT often fails to improve the I/O performance of the compute nodes~\cite{dftm}. Another category of methods for reliable cloud computing applies a Markov based model of the system~\cite{luo2019improving} to examine the potential failures and take quick recovery measures. As these methods are applicable in edge scenarios, we consider these as baselines in our experiments. 

\textbf{AI Methods.} Recently, many unsupervised methods have started utilizing AI techniques for fault prevention and fault recovery. The Energy-efficient Checkpointing and Load Balancing (ECLB)~\cite{eclb} approach applies Bayesian methods and neural networks to classify host machines into overloaded, under-loaded and normal categories. This is done using heuristic thresholds instead of supervised class labels. It then applies this classification to select tasks from the hosts in the overloaded category. The target hosts are selected from the under-loaded category. This approach primarily relies on computational load-balancing for fault recovery, but does not consider other resource types like memory, disk and network. Very recent works also propose a few-shot learning method for fault detection~\cite{won2021performance}. Such methods train using supervised labels and do not present a mechanism to recover from faults once detected, and hence are not considered baselines in our work. Other recent methods utilize deep neural networks to execute fuzzy clustering~\cite{awgg, li2018comparison}. For instance, the Adaptive Weighted Gath-Geva (AWGG)~\cite{awgg} clustering method is an unsupervised model that detects faults using stacked sparse autoencoders to reduce detection times. Due to the better relative performance than prior works, the ECLB and AWGG methods are also considered baselines in our experiments. Another recent class of methods applies neural networks to reconstruct the last state of the system~\cite{audibert2020usad, topomad}. The reconstruction error is used as an indicator of the likelihood of the current state being anomalous. For instance, TopoMAD~\cite{topomad} utilizes a topology-aware neural network that is composed of a Long-Short-Term-Memory (LSTM) and a variational autoencoder (VAE) to detect faults. However, the reconstruction error is only obtained for the latest state, which limits them to use reactive fault recovery policies. Moreover, TopoMAD is not agnostic to the number of hosts or workloads as it assumes a maximum limit of the active tasks in the system. Nevertheless, our experiments show that TopoMAD outperforms other baselines and is considered as a benchmark in our evaluation.

\section{Methodology}
\label{sec:method}

\begin{figure}
    \centering \setlength\abovecaptionskip{-0.1\baselineskip}
    \includegraphics[width=0.75\linewidth]{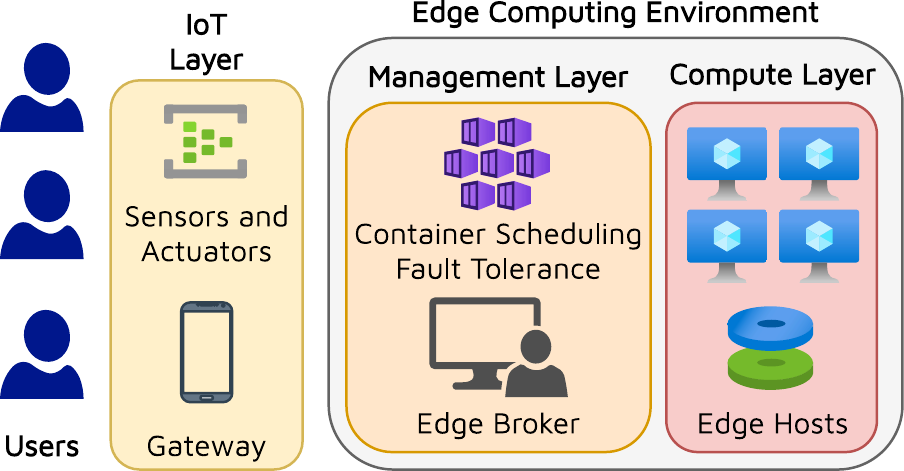}
    \caption{System Model}
    \label{fig:system}
\end{figure}

\begin{figure*}[] 
    \centering \setlength\abovecaptionskip{-0.1\baselineskip} \setlength\belowcaptionskip{-0.5\baselineskip}
    \includegraphics[width=\linewidth]{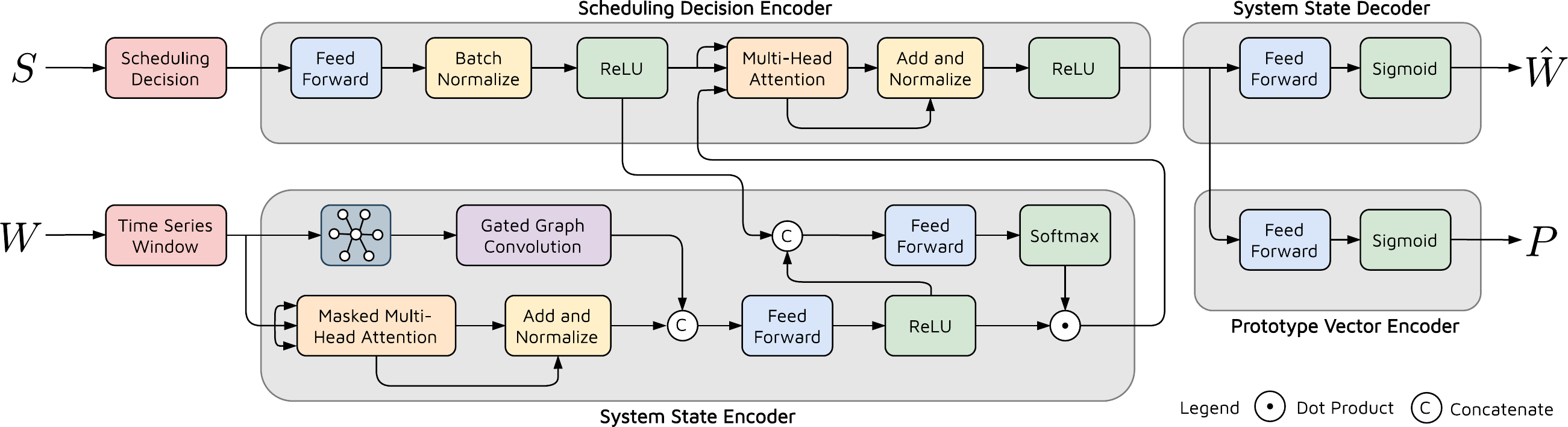}
    \caption{The DeepFT Surrogate Model.}
    \label{fig:model}
\end{figure*}

\subsection{Environment Assumptions and Problem Formulation}

\textbf{System Model.} We assume a standard edge computing environment with heterogeneous nodes in a master-slave fashion as summarized in Figure~\ref{fig:system}: 
\begin{itemize}[leftmargin=*]
    \item The {\it Compute Layer} consists of a fixed set of $m$
%multiple unreliable and resource constrained
edge hosts (worker nodes), denoted as $H = \{h_1, \ldots, h_m\}$. 

\item The {\it Management Layer} \blue{receives all incoming tasks} and includes a \blue{broker} node with sufficient compute capacity to perform fault-tolerant tasks scheduling.
\end{itemize}
We consider that an execution in this environment runs for a bounded time and that the timeline is divided into fixed-sized scheduling intervals, where $I_t$ denotes the $t$-th interval ($t$ ranging from 0 to $T$). The edge broker can sample the resource utilization metrics of all hosts at any time (e.g., CPU, RAM, disk/network bandwidth, some additional fault-related metrics including consumption of the swap space, disk buffers, network buffers, disk and network I/O waits~\cite{tanenbaum1997operating}).

\textbf{Fault Model.} We consider resource contention type faults, which frequently occur in edge workers, while executing data-intensive applications. Here, a worker node may become unresponsive due to resource over-utilization~\cite{vasilakos2020towards}, \blue{we thus consider faults of the form of CPU, RAM or Disk contention that we raised when their utilization is greater than dynamically set thresholds}. As in prior work, edge hosts are connected to the same power supply and unrecoverable faults like outages are ignored~\cite{dftm, javed2018cefiot, bagchi2019dependability}. We aim to prevent over-utilization faults by adjusting the task scheduling and migration decisions.

\textbf{Workload Model.} We assume a bag-of-tasks workload model, where a set of independent tasks enter the system at the start of each scheduling interval. These are generated from the users and are transferred to the edge broker via gateway devices or IoT sensors. To evaluate the proposed method in a controlled environment, we abstract out the users and IoT layers in our experiments and apply a discrete probability distribution to realize tasks as container instances (see Section~\ref{sec:experiments}). Each task has an associated SLO deadline. 

\textbf{Formulation.} Formally, at the start of the interval $I_t$, the edge broker needs to make a decision denoted as $S_t$, that is a mapping of new tasks in the interval $I_t$ as well as the active tasks from interval $I_{t-1}$ to hosts $H = \{h_1, \ldots, h_m\}$.

\begin{table}[t]
    \centering
    \caption{Symbol Table}
    \label{tab:symbols}
    \resizebox{\linewidth}{!}{
    \begin{tabular}{@{}ll@{}}
    \toprule 
    Symbol & Meaning\tabularnewline
    \midrule
    $H$ & Set of hosts in the system\tabularnewline
    $h_i$ & $i$-th host in $H$ such that $i \in \{1, \ldots, m\}$\tabularnewline
    $I_t$ & $t$-th interval\tabularnewline
    $x_t$ & System state at the start of $I_t$\tabularnewline
    $W_t$ & Sliding window at $I_t$ of system states of length $k$ \tabularnewline
    $x_i$ & $i$-th state in $W_t$ such that $i \in \{t-k+1, \ldots, t\}$\tabularnewline
    $\hat{W}_t$ & Predicted reconstruction of $W_t$ \tabularnewline
    $S_t$ & Scheduling decision in $I_t$ \tabularnewline
    $f_t$ & Fault score as the square of the L2 distance between $W_t$ and $\hat{W}_t$ \tabularnewline 
    $P_t$ & Prototype vector for fault classification\tabularnewline
    $c_i$ & $i$-th class in the set of faults $i \in \{1, \ldots, j\}$\tabularnewline 
    $c_0$ & No fault class or No-Anomaly-Prototype (NAP) class \tabularnewline 
    \bottomrule
    \end{tabular}}
\end{table}

For new tasks, this becomes a scheduling decision and for existing active tasks, it acts as a preemptive migration decision (if the target host is different from the current host). Now, at the beginning of interval $I_t$, for an input time-series of system states $\{x_0, \ldots, x_{t}\}$, the broker needs to predict the next system state, \textit{i.e.}, $x_{t+1}$. Here, the system state ($x_t$) consists of values for an arbitrary set of resource usage metrics for all active tasks and hosts at the start of $I_t$. Instead of directly using the system states, we consider a sliding window of length $k$ to capture the temporal contextual trends:
\[W_t = \{x_{t-k+1}, \ldots, x_{t}\}.\]
We use replication padding~\cite{liu2018partial} for the first $k$ intervals to ensure the time-series window is always of the same size. Also, to ensure robust predictions in the DeepFT model, we normalize these time-series windows using min-max scaling to be in the range $[0,1]$. \blue{Table~\ref{tab:symbols} summarizes all symbols.}

Using the input window $W_t$ and an input scheduling decision $S_t$, the model needs to predict whether there is likely to be a fault in the next system state, \textit{i.e.}, $x_{t+1}$. In lieu of directly \blue{predicting whether the next system state is faulty}, we predict a fault score $f_t$ using a predicted reconstruction of the next window $W_{t+1}$, denoted as $\hat{W}_{t+1}$. The fault score $f_t$ is obtained by calculating the deviation between the true window $W_{t+1}$ and its reconstruction $\hat{W}_{t+1}$. We also predict the type of fault (CPU/RAM/Disk contention), referred to as the fault class, of the next state in the form of a prototype embedding $P_{t+1}$. A prototype embedding~\cite{snell2017prototypical} is a dense vector representation of the fault class a particular input belongs to. Without loss in generality, whenever unambiguous, we drop the subscripts for simplicity. Hence, we only refer to the inputs and outputs as $W$, $S$, $\hat{W}$,  $P$ and $f$.

\subsection{DeepFT Surrogate Model}
\label{sec:deepft}

% key insights and reasons why we do what we do
As discussed in the previous subsection, for an input system sliding window $W$ and scheduling decision $S$, the surrogate model of DeepFT predicts a future system state $\hat{W}$ and fault classification prototype vector $P$. The surrogate model used in DeepFT is a composite neural network presented in Figure~\ref{fig:model}. The DeepFT model has four main components: (1) Scheduling Decision Encoder, (2) System State Encoder, (3) System State Decoder and (4) Prototype Vector Encoder. 

% matrix/tensor forms of inputs 
Considering we have $p$ tasks in the system as a sum of new and active tasks, the scheduling decisions of these are encoded as one-hot vectors of size $m$ (number of hosts). We thus get a matrix of scheduling decisions ($S$) of size $[p \times m]$. Also, the $n$ features of $m$ hosts and $p$ tasks in the form of resource utilization metrics, each state window ($W$) is encoded as a $[(m+p) \times n \times k]$ tensor. All operations are performed in a factored fashion, \textit{i.e.}, the neural network operates on $S$ as a batch of $p$ vectors, each having dimension $m$, and on $W$ as a batch of size $m+p$ tensors of size $n\times k$.

% different components working
\textbf{Scheduling Decision Encoder.} The decision encoder encapsulates the scheduling decision in the form of a compressed vector representation. This representation uses an encoding of the time-series window ($E^W$ generated by the state encoder) as an input to facilitate the prediction of the next window and the fault class prototype embeddings. To do this, we utilize a feed-forward layer to bring down the dimension size of the input with batch-normalization and rectified linear unit ($\mathrm{ReLU}$) activation. \blue{The $\mathrm{ReLU}$  is a non-linear activation that outputs the input directly if it is positive and zero otherwise. Such non-linear activation functions allow the neural network to learn complex data patterns and $\mathrm{ReLU}$ facilitates faster training~\cite{aima}. Thus,}
\begin{equation}
    E^S = \mathrm{ReLU}(\mathrm{BatchNorm}(\mathrm{FeedForward}(S))).
\end{equation}
\blue{Here, a feed-forward layer performs a matrix-multiplication operation with a tunable set of parameters. A batch-normalization layer normalizes the input data across each dimension to zero mean and unit variance, which ensures faster and more robust training~\cite{ioffe2015batch}. }

We then use a multi-head self-attention operation~\cite{vaswani2017attention} with the window encoding $E^W$ to provide a representation that focuses on only a subset of the tasks and hosts that may result in faults, relieving the downstream predictors from inferring over tasks/hosts that have low fault probability. \blue{Self-attention passes the input through a feed-forward layer to generate a set of weights, which are then used to take a convex combination of the encodings of multiple tasks in $E^W$. In the multi-headed version, the input is passed through multiple such feed-forward layers to capture the correlations across multiple such task sets.} This self-attention technique allows {\it the designed model to scale efficiently} with the number of hosts or active tasks in the system (demonstrated with a visualization in Section~\ref{sec:training}). This with the layer-normalization operation and the $\mathrm{ReLU}$ activation gives
\begin{align}
\begin{split}
    E^S_1 &= \mathrm{Norm}(E^S\! +\! \mathrm{MultiHeadAtt}(E^S, E^S, E^W)),\\[-2pt]
    E^S_2 &= \mathrm{ReLU}(E^S_1).
\end{split}
\end{align}

\textbf{System State Encoder.} The state encoder creates a succinct encoding of the system state time-series window, conditioned on the task scheduling decision to facilitate the prediction of the next system state window.  We first form a graph using the task schedule $S$, such that there is an edge from host $h_i$ to host $h_j$ if there is a task migration from $h_i$ to $h_j$ in schedule $S$. The $n$ characteristics of each host in $x_t$ are then used to populate the feature vectors of the nodes in the graph. We denote the feature vector of host $h_i$ as $e_i$. We then pass the graph through a gated-graph convolution network to capture the inter-host dependencies rising from the new task allocation $S$. Here, the features for host $h_i$ are aggregated over one-step connected neighbors $n(i)$ in the graph over $r$ convolutions, resulting in an embedding $e^{r}_{i}$ for each host node in the graph. Specifically, the gating stage is realized as a Gated Recurrent Unit (GRU) resulting in \emph{graph-to-graph} updates~\cite{ruiz2020gated} as: 
\begin{align}
\begin{split}
    e_i^{0} &= \mathrm{tanh} ( W\ e_{i} + b ),\\[-2pt]
    x^q_i &= \sum_{j \in n(i)} W^q e_{j}^{q-1} ,\\[-2pt]
    e^q_{i} &= \mathrm{GRU} ( e^{q-1}_i, x^{q}_{i} ),
\end{split}
\end{align}
\blue{where the second equation performs the convolutions of the features of immediate neighbors in the graph. However, for large-scale graphs, to ensure that we capture the inter task and host correlations, we perform the above convolution step $r$ times. Here, a GRU is a recurrent neural network that decides the weightage of the output of the previous convolution iteration with respect to the latest iteration. This allows the model to efficiently scale with the size of input graph without significantly losing performance.}
The stacked representation for all hosts is represented as $E^H$. We also pass the time-series window through a multi-head self attention network. 
\begin{align}
\begin{split}
    E^W_1 &=  \mathrm{Mask}(\mathrm{MultiHeadAtt}(E^W, E^W, E^W)),\\[-2pt]
    E^W_2 &= \mathrm{Norm}(W + E^W).
\end{split}
\end{align}
The first attention operation is masked to prevent the encoder from looking at the datapoints for future timestamp values at the time of training as all time-series windows are given at once to allow parallel training. The resulting window encoding is denoted as $E^W_2$. We then apply dot-product attention~\cite{bahdanau2015neural} conditioned on the decision encoding $E^S$, giving the final window encoding as
\begin{align}
\begin{split}
    E^W_3 &= \mathrm{ReLU}(\mathrm{FeedForward}([E^W_2, E^H])),\\[-2pt]
    attn &=  \mathrm{softmax}(\mathrm{FeedForward}([E^W_2, E^S]),\\[-2pt]
    E^W &=  attn \cdot E^W_3.
\end{split}
\end{align}
\blue{This attention is similar to the multi-headed attention operation without multiple heads and also uses $E^S$ as the input of the feed-forward layer, allowing us to generate an output conditioned on the input scheduling decision $S$.}

\textbf{System State Decoder.}
Now that we have the embedding that captures both decision and state $E^S_2$, we generate the reconstructed window as
\begin{equation}
    \hat{W} = \mathrm{sigmoid}(\mathrm{FeedForward}(E^S_2)).
\end{equation}
Here, the sigmoid function allows us to bring the output in the range $[0, 1]$, the same as that of the normalized true window ($W$). To generate the fault score we only consider upward spikes of the true window from the reconstructed window. This is due to the nature of our state data, \textit{i.e}, resource utilization metrics leading to faults only when there is a sudden increase in CPU/RAM/disk/network consumption. The $\mathrm{ReLU}$ activation is apt for this as it gives a zero fault score when the true utilization metrics are lower than the predicted ones.  Moreover, the fault detection label is decided by comparing the fault score with a threshold. We use Peak-Over-Threshold (POT)~\cite{siffer2017anomaly} to generate the fault label for input $W$. POT is a dynamic thresholding technique that uses extreme value theory to set thresholds for each input dimension. Thus, our fault score and label \blue{(1 if faulty and 0 otherwise)} are calculated as
\begin{align}
\begin{split}
    \label{eq:fault_label}
    f &= \norm{\mathrm{ReLU}(W - \hat{W})}^2,\\
    l &= \mathds{1}(f > POT(W)).
\end{split}
\end{align}

\textbf{Prototype Vector Encoder.} Our prototype encoder is motivated by prototypical networks proposed in prior work~\cite{snell2017prototypical}. Such networks are standard few-shot learning models for supervised clustering of the input data. Few-shot learning is a paradigm in which models aim to achieve the required task using only a few data points. This allows the DeepFT model to adapt to changing environments in a few scheduling intervals. Prototypical networks perform clustering by maintaining representative prototype vectors for each class. Such networks generate a prototype embedding for an input and ensure that the output vector is close to the prototypes for the correct class and far from the prototypes of the incorrect classes. 
The prototype embedding is generated from the $E^S_2$ embedding as
\begin{equation}
    P = \mathrm{sigmoid}(\mathrm{FeedForward}(E^S_2)).
\end{equation}

\subsection{Offline Model Training}
\label{sec:offline_training}
\begin{algorithm}[t]
    \begin{algorithmic}[1]
    \Require
    \Statex Deep Surrogate Model $M$
    \Statex Dataset used for training $\{W_t, S_t\}_{t=1}^T$
    \Statex Iteration limit $L$
    \State Initialize weights in $M$. Set $l \gets 0$
    \State \hspace{1em} \textbf{do}
    \State \hspace{2em} Calculate $(\mu_i, \sigma_i)$ for each class $\{c_i\}_{i=0}^j$ \label{line:clusterids}
    \State \hspace{2em} \textbf{for} ($t = 1 \text{ to } T$) \textbf{do}
    \State \hspace{3em} $\hat{W}_t, P_t \gets M(W_t, S_t)$ \label{line:inference}
    \State \hspace{3em} Calculate $l_t$ using~\eqref{eq:fault_label} 
    \State \hspace{3em} \textbf{if} ($l_t$) \textbf{do}
    \State \hspace{4em} $\phi = \argmin_{i=1}^j D(P_t, c_i)$ \label{line:cj}
    \State \hspace{3em} \textbf{else}
    \State \hspace{4em} $\phi = 0$ \label{line:c0}
    \State \hspace{3em} $L_R = \norm{\hat{W}_t - W_{t+1}}^2$ \label{line:l1}
    \State \hspace{3em} $L_T = D(P_t, c_\phi)  -  \sum_{i \neq \phi} D(P_t, c_i)$ \label{line:l2}
    \State \hspace{3em} Update weights of $M$ using $L_R + L_T$ 
    \State \hspace{3em} $l \gets l + 1$
    \State \hspace{1em} \textbf{while} $l < L$
    \State \textbf{return} trained $M$
    \end{algorithmic}
\caption{The DeepFT offline training algorithm}
\label{alg:training}
\end{algorithm}

\label{sec:training}
We now describe the training process for the DeepFT surrogate model, summarized in Algorithm~\ref{alg:training}. To do this, we collect a dataset of scheduling decisions and time-series window pairs $\{W_t, S_t\}_{t=1}^T$. To do this, we use a random scheduler to cover as much of the decision space as we can. We use the model to find the prototype vectors of each data pair and calculate their mean and standard deviations for each class $c_i$ as $\{(\mu_i, \sigma_i)\}_{i=0}^j$ \blue{(line~\ref{line:clusterids})}. For an input pair $(W_t, S_t)$, the model generates an output $(\hat{W}_t, P_t)$ (line~\ref{line:inference}).  Now, for the state decoder, we define the reconstruction loss as the mean-square-error 
\begin{equation}
\label{eq:loss_r}
    L_R = \norm{\hat{W}_t - W_{t+1}}^2.
\end{equation}
\blue{Minimizing this loss ensures that our model generates a precise prediction of the next state.}

\begin{figure}
    \centering \setlength\abovecaptionskip{-0.2\baselineskip}
    \includegraphics[width=0.95\linewidth]{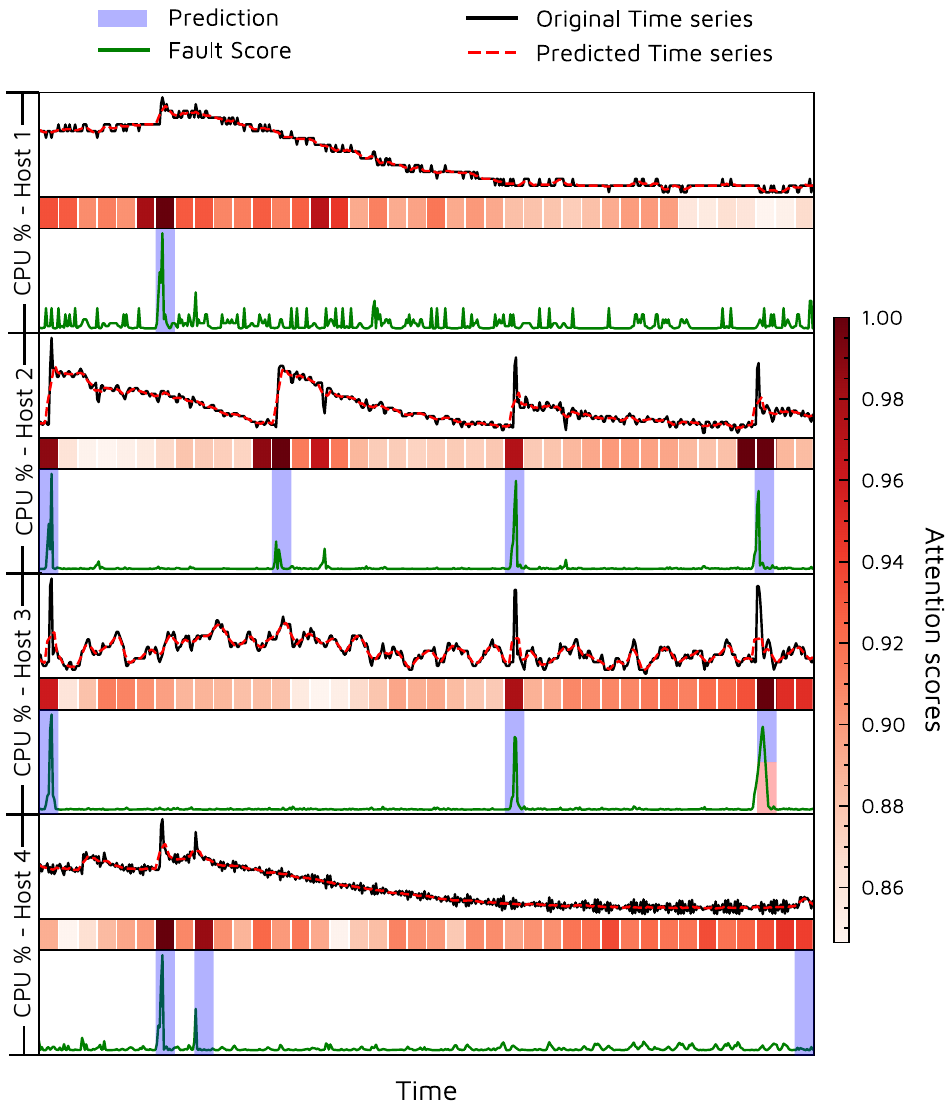}
    \caption{ Visualization of attention weights, truncated with the CPU utilization of four hosts in the system. The attention operation allows the model to scale with the number of hosts or workloads.}
    \label{fig:attention}
\end{figure}

The prototype encoder outputs the embedding $P_t$. In our case, true class labels are \textit{a-priori} unknown. To circumvent this issue, we use the fault prediction from the state decoder. \blue{Thus, we first generate the fault prediction label $l_t$ using~\eqref{eq:fault_label}.} If a faulty state is predicted, we label it as one of the $j$ classes $\{c_1, \ldots, c_j\}$. Here, $j$ is a hyperparameter corresponding to the different types of faults in an edge worker. Motivated by~\cite{medina2020selfsupervised}, to generate the ground-truth class in the case that an anomaly is detected by the state decoder, we choose the class corresponding to the closest class prototype from the output $P_t$ (line~\ref{line:cj}). This class label may not correspond to any semantically defined fault, but helps the model implicitly segregate faults and improve learnability~\cite{medina2020selfsupervised}. If there is no fault, we call it the No-Anomaly-Prototype (NAP) class, denoted as $c_0$ (line~\ref{line:c0}). \blue{This training style requires only a few datapoints to generate a class prototype as the centroid of the points in that class; hence, it is called few-shot learning.}  

As the fault distribution is highly skewed in typical executions with most states belonging to the NAP class, we use the Kullback–Leibler divergence between the output $P_t$ and a class representative pair $(\mu_i, \sigma_i)$:
\begin{equation}
\label{eq:aleatoric}
D(P_t, c_i) = \frac{(\mu_i - P_t)^2}{2 \sigma_i^2} + \frac{1}{2}\ln{\sigma_i^2}.
\end{equation}
\blue{Unlike the Euclidean distance between the predicted prototype $P_t$ and the mean of the class $\mu_i$, this divergence accounts for the overall distribution of the class vectors.} The ground-truth class ($c_\phi$) is then determined as
\begin{equation}
    c_\phi\ \text{s.t.}\ \phi = \left\{
  \begin{array}{@{}ll@{}}
    \argmin_{i=1}^j D(P_t, c_i), & \text{if}\ l_t \\
    0, & \text{otherwise}.
  \end{array}\right.
\end{equation}

To train the model, we apply the triplet loss~\cite{snell2017prototypical} to ensure the predicted prototype embedding is closer to the true class $c_\phi$ \blue{(minimizing $D(P_t, c_\phi)$)} and far from the incorrect class vectors $c_i \forall i \neq \phi$ \blue{(maximizing $\sum_{i \neq \phi} D(P_t, c_i)$)}:
\begin{equation}
\label{eq:loss_t}
    L_T = D(P_t, c_\phi)  -  \sum_{i \neq \phi} D(P_t, c_i).
\end{equation}

\textbf{Visualization of attention weights.} Figure~\ref{fig:attention} visualizes the attention weights averaged across the heads of the attention operation in the scheduling decision encoder. The model is trained using a dataset collected from a physical setup (see details in Section~\ref{sec:experiments}) using a random scheduler. We show the ground-truth (in black) and predicted time-series (in red) using the first state vector of the reconstructed window. The fault scores are also shown in green. The red heatmap shows the average attention weights for each window. The attention operation compresses the scheduling decision matrix to a fixed size vector allowing the model to scale to higher number of hosts or workloads in the system. For instance, the number of parameters of the model increases by only 3\% or 1\% for each new task or host in the system. Moreover, there is a high correlation between the fault scores and attention weights, showing how attention facilitates focusing on only those dimensions that may result in faults. The fault labels are also highlighted using the POT thresholds.

\begin{algorithm}[t]
    \begin{algorithmic}[1]
    \Require
    \Statex Trained Deep Surrogate Model $M$
    \Statex Use converged $\{(\mu_i, \sigma_i)\}_{i=0}^j$ from training
    \State \textbf{for}($t = 1 \text{ to } T$)
    \State \hspace{1em} $S_t \gets S_{t-1}$ \textbf{if} $t>0$ \textbf{else} initialize randomly \label{line:start}
    \State \hspace{1em} $\hat{W}_t, P_t \gets M(W_t, S_t)$
    \State \hspace{1em} $W_{t+1} \gets \mathrm{Sim}(W_t, S_t)$ \label{line:sim}
    \State \hspace{1em} $L_O = \norm{\mathrm{ReLU}(W_{t+1} - \hat{W}_t)}^2 + D(P_t, c_0)$ \label{line:lo}
    \State \hspace{1em} Optimize $S_t$ using $L_O$ and $\mathrm{Adam \hspace{0.5em}optimizer}$
    \State \hspace{1em} Update weights of $M$ using $L_R + L_T$ in Eqs.\eqref{eq:loss_r},\eqref{eq:loss_t} \label{line:tune}
    \State \hspace{1em} \textbf{return} $S$ \label{line:s}
    \end{algorithmic} 
\caption{The DeepFT scheduling algorithm}
\label{alg:testing}
\end{algorithm}

\subsection{Scheduling and Online Model Fine-Tuning}
\label{sec:online_training}
We now describe how we use the surrogate model to generate scheduling decisions for fault-tolerant computing, summarized in Algorithm~\ref{alg:testing}. Our method is motivated by gradient-based optimization using backpropagation of the output of a surrogate model to the input, called the GOBI method~\cite{tuli2021cosco}. At the beginning of interval $I_t$, we start from $S_{t-1}$ (if $t>0$ otherwise initialize randomly) and optimize it using DeepFT's surrogate model, denoted as $M$ (line~\ref{line:start}). Compared to initializing $S_t$ randomly, this helps us to reduce the number of migrations. Also, \blue{to generate our self-supervision next-state $W_{t+1}$ at test time,} we use a co-simulator to find the next state using the current state and scheduling decision, denoted as $W_{t+1} \gets \mathrm{Sim}(W_t, S_t)$ (line~\ref{line:sim}). \blue{Our co-simulator runs a single-step execution trace of the scheduling decision $S_t$ using the workload characteristics $W_t$ to estimate the system state at the start of the next interval, \textit{i.e.}, $W_{t+1}$.} For input $(W_t, S_t)$ and model output $\hat{W}_t, P_t \gets M(W_t, S_t)$, we define optimization loss as
\begin{equation}
    L_O = \norm{\mathrm{ReLU}(W_{t+1} - \hat{W}_t)}^2 + D(P_t, c_0).
\end{equation}
Optimizing using this loss minimizes the fault score and updates the prototype vector to move towards the NAP class. As we update the scheduling decisions to minimize $L_O$, when converged, we hypothesize that this decision should facilitate fault avoidance. To optimize the scheduling decision $S_t$, we can use the stochastic gradient descent
\begin{equation}
    S_t \gets S_t - \nabla_{S_t} L_O.
\end{equation}
However, advances like root-mean-square propagation and adaptive gradients facilitate faster convergence; hence, we utilize the $\mathrm{Adam}$ optimizer~\cite{kingma2014adam} with cosine annealing~\cite{loshchilov2016sgdr} to optimize the scheduling decision. \blue{Thus, running an optimization over the scheduling decision $S_t$ to minimize $L_O$ should, in principle, give us a scheduling decision, say $S^*_t$, such that $P_t$ with $S^*_t$ as input belongs to the NAP class and hence, it does not lead to a contention-like fault in the subsequent interval.}  The final converged decision is applied to schedule and migrate tasks. As we do not need the ground-truth labels, we can fine-tune the surrogate model on-the-fly at test time (line~\ref{line:tune} of Alg.~\ref{alg:testing}). 

\section{Experiments}
\label{sec:experiments}
\noindent
We compare the DeepFT method against the baselines DFTM~\cite{dftm}, ECLB~\cite{eclb}, PCFT~\cite{pcft}, AWGG~\cite{awgg} and TopoMAD~\cite{topomad} (more details in Section~\ref{sec:related_work}). As TopoMAD and AWGG are only fault-detection methods, we supplement them with the PSO based preemptive migration scheme for fault-recovery from the next best baseline, \textit{i.e.}, PCFT. We use hyperparameters of the baseline models as presented in their respective papers. We train all deep learning models using the PyTorch-1.8.0~\cite{paszke2019pytorch} library. 

\begin{figure}[t]
    \centering \setlength\abovecaptionskip{-0.1\baselineskip}
    \includegraphics[width=0.4\linewidth]{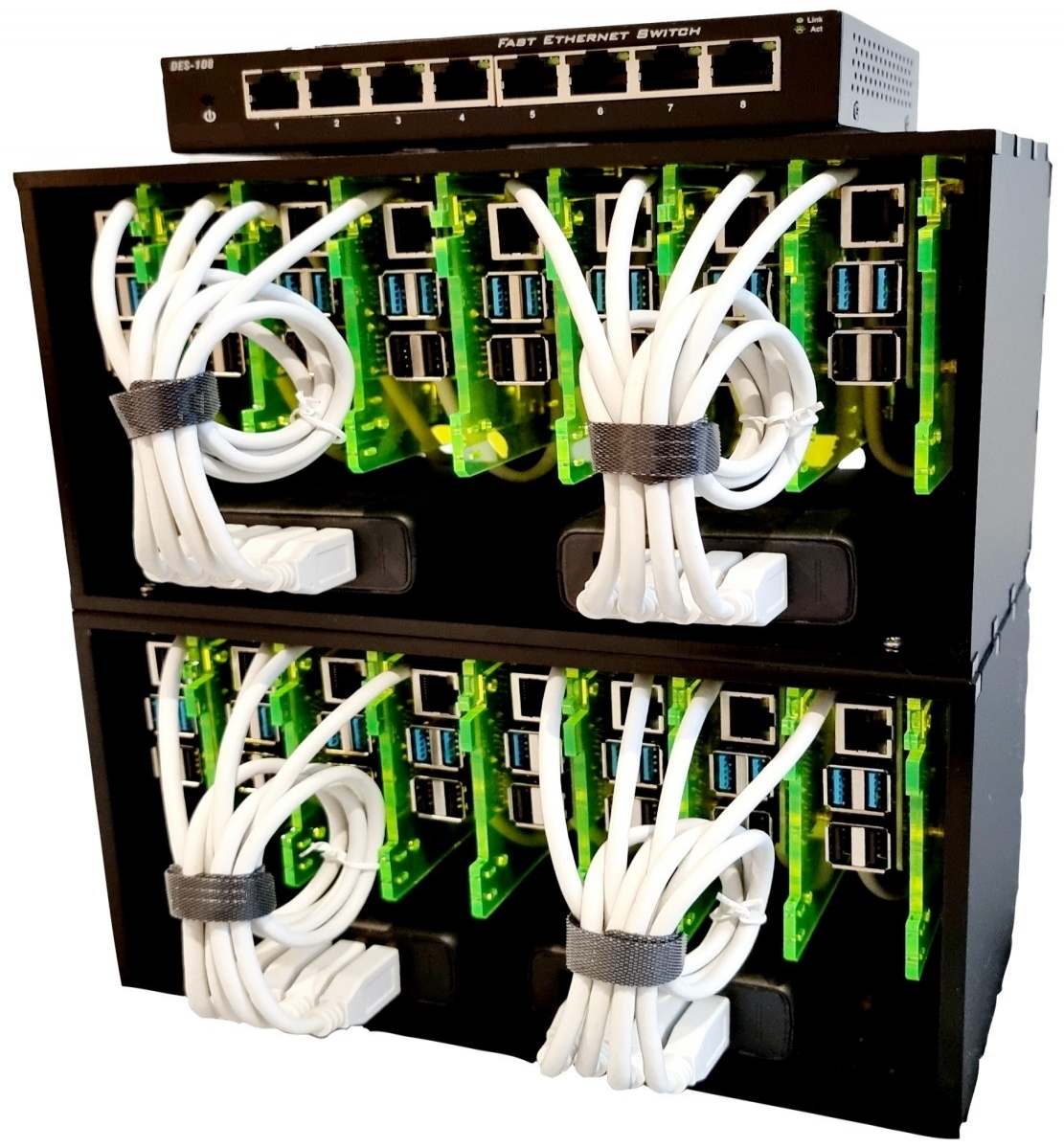}
    \caption{Raspberry Pi Cluster as evaluation platform.}
    \label{fig:rpi}
\end{figure}

\subsection{Evaluation Setup}
As our experimental setup, we use an edge computing cluster of 16 Raspberry Pi 4B nodes as shown in Figure~\ref{fig:rpi}.
This cluster consists of eight 4-GB RAM nodes and another eight 8-GB RAM nodes. Instead of using the watt-meter, we consider the power consumption models from the commonly-used Standard Performance Evaluation Corporation (SPEC) benchmarks repository~\cite{spec}. We run all experiments for 100 scheduling intervals, with each interval being 5 minutes long, giving a total experiment time of 8 hours 20 minutes. We average over 5 runs and apply diverse workload types to ensure statistical significance. The offline model training (explained in Section~\ref{sec:offline_training}) was performed on a system with configuration: Intel i7-10700K CPU, 64GB RAM, Nvidia RTX 3080 and Windows 11 OS. The scheduling and the online fine-tuning of the model was performed by the edge broker with configuration: laptop with Intel i3-1115G4 and 8GB RAM. 

To generate the tasks in our system, we utilize the \textit{DeFog} applications~\cite{mcchesney2019defog}. DeFog is a fog computing benchmark suite that consists of various real-world application instances used in edge and cloud computing environments~\cite{tuli2021pregan}. The three specific application types: Yolo, PocketSphinx and Aeneas were used in our experiments due to their volatile characteristics and heterogeneous resource requirements. At the start of each scheduling interval, we create $Poisson(\lambda)$ new tasks, sampled uniformly from the three applications. Poisson distribution is a natural choice for a bag-of-tasks workload model, common in edge environments~\cite{mao2016dynamic, basu2019learn, tuli2021cosco}. Our tasks are executed using Docker containers~\cite{boettiger2015introduction}, common in edge platforms~\cite{morabito2017evaluating}.

\subsection{Evaluation Metrics}
\label{sec:evaluation_metrics}

To collect the training data for the surrogate model, we use a random scheduler. To test the fault detection and diagnosis performance, we utilize historical log data collected by running the GOBI approach~\cite{tuli2021cosco} on our testbed. To generate the ground-truth fault labels, we consider the Anomaly Detection Engine for Linux Logs (ADE) tool~\cite{ade}. For every state of the dataset, we use the fault flag from the ADE tool for the subsequent state as our ground-truth label. This fault flag is raised when one or more of the metrics like CPU, RAM, disk and network utilization is/are greater than dynamically set thresholds. These may be caused due to memory leaks, buffer overloads and resource throttling.

To evaluate fault-detection efficacy, we utilize commonly considered metrics including accuracy, precision, recall and F1 score. For fault-diagnosis (\textit{i.e.}, the prediction of the specific hosts that have faults), the factored style prediction of fault scores and labels enables us to output labels for each host independently. We use two popular metrics for comparison: (1) $\mathrm{HitRate@100\%}$ is the measure of how many ground truth dimensions have been included in the top candidates predicted by the model~\cite{omnianomaly}, (2) Normalized Discounted Cumulative Gain ($\mathrm{NDCG@100\%}$)~\cite{jarvelin2002cumulated}. To also compare how well the model is performing compared to our reference scheduler GOBI that is used to collect our test dataset, we present a more insightful metric called the "Improvement Ratio". For any model $X$, we calculate this using the co-simulator to obtain the QoS estimate (denoted as $\mathrm{QoS}(.)$) of the scheduling decision of $X$ ($S_t^X$) and that of GOBI ($S_t^{GOBI}$). Thus,
\begin{equation}
    \text{Impr. Ratio} = \frac{1}{T} \sum_{t = 1}^T \mathds{1}{( \mathrm{QoS}(S_t^X) > \mathrm{QoS}(S_t^{GOBI}) )}.
\end{equation}
This denotes the ratio of the times the model $X$ can predict a better scheduling decision than the reference. As GOBI is oblivious to future faults, this metric should always be $>0.5$. The QoS estimate is calculated using a combination of metrics like energy consumption and response time, common in edge computing deployments~\cite{tuli2021cosco, basu2019learn}. Thus,
\begin{equation}
    \mathrm{QoS(S_t)} = 1 - \alpha \cdot ART_t - \beta \cdot AEC_t.
\end{equation}
Here, $ART_t$ is the average response time of the tasks leaving the system at the end of $I_t$ and $AEC_t$ is the average energy consumption of the system in $I_t$. Both are obtained by executing $S_t$ on a co-simulator of a scheduling interval.  Also, $\alpha, \beta$ are convex combination weights that can be set as per user requirements. For our experiments, we consider $\alpha = \beta = 0.5$ as per prior work~\cite{tuli2021cosco}. Note that a higher QoS score is better.

To compare the scheduling times, we normalize them with respect to the scheduling time of the reference GOBI scheduler and call these overhead ratios. We also consider standard QoS metrics, including resource utilization, fraction of service level objective (SLO) violations, fairness and average migration time. We consider the relative definition of SLO (as in~\cite{tuli2021cosco}) where the deadline is the 90$^{th}$ percentile response time for the same application (Yolo/Pocketsphinx/Aeneas) on the state-of-the-art baseline TopoMAD. For fairness, we use the Jain's fairness index~\cite{tuli2021cosco}. Migration time is the time spent to live-migrate the running container instances in the system.

\begin{figure}[t]
    \centering  \setlength\belowcaptionskip{-0.2\baselineskip}
    \subfigure[Loss]{
    \includegraphics[width=0.48\linewidth]{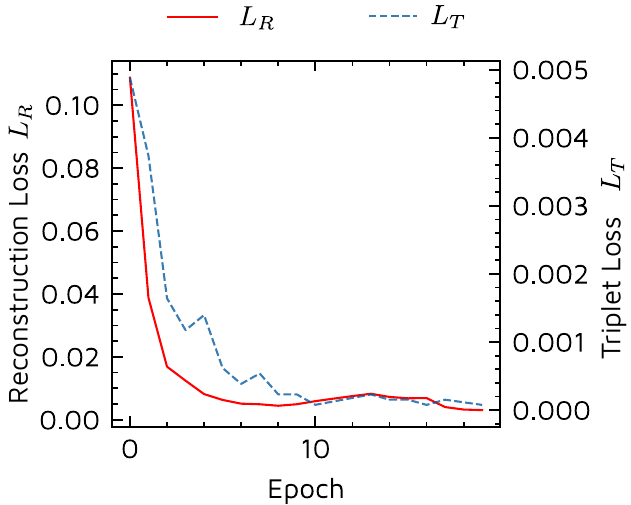}
    \label{fig:loss}
    }%
    \subfigure[Prediction Accuracy]{
    \includegraphics[width=0.49\linewidth]{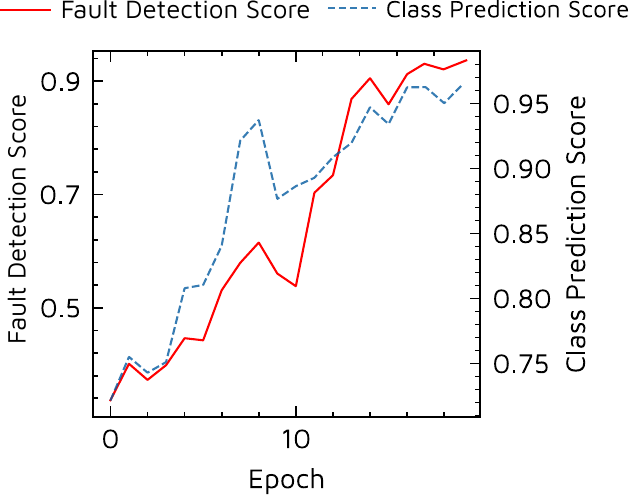}
    \label{fig:score}
    }
    \caption{Convergence plots for the DeepFT model.}
    \label{fig:convergence}
\end{figure}

\begin{figure}[!t]
    \centering \setlength\abovecaptionskip{-0.2\baselineskip}
    \includegraphics[width=\linewidth]{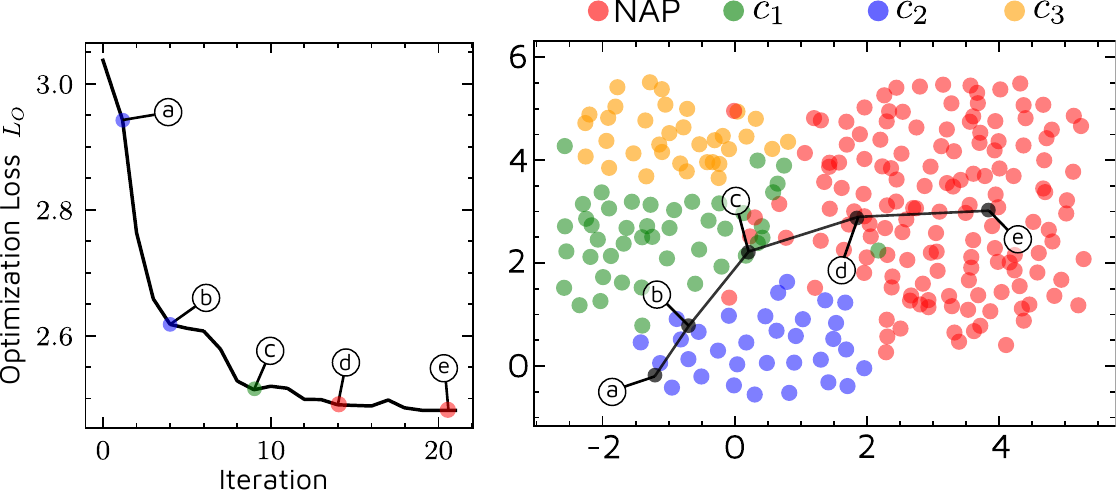}
    \caption{\footnotesize Visualization of the optimization loop for a sample scheduling decision. The plot (left) shows how the optimization loss $L_O$ reduces with iteration count. The t-SNE plot (right) shows how the prototype vector of the decision changes and converges to be within the NAP (no-fault) class.}
    \label{fig:scores}
\end{figure}

\begin{table*}[!t]
    \centering
    \caption{Performance scores of DeepFT and the baseline methods with standard deviation. The best scores are shown in bold.}
    \resizebox{\textwidth}{!}{
    \begin{tabular}{@{}lcccccccc@{}}
    \toprule 
    \multirow{2}{*}{Method} & \multicolumn{4}{c}{Detection} & \multicolumn{2}{c}{Diagnosis} & \multirow{2}{*}{\begin{tabular}{@{}c@{}}Overhead \\ Ratio\end{tabular}} & \multirow{2}{*}{\begin{tabular}{@{}c@{}}Improvement \\ Ratio\end{tabular}}\tabularnewline
    \cmidrule{2-7}
     & Accuracy & Precision & Recall & F1 Score & HR@100 & NDCG@100 & & \tabularnewline
     \midrule
    DFTM~\cite{dftm} & 0.8731 \scriptsize{$\pm$0.0234} & 0.7713 \scriptsize{$\pm$0.0823} & 0.8427 \scriptsize{$\pm$0.0199} & 0.8054 \scriptsize{$\pm$0.0872} & 0.5129 \scriptsize{$\pm$0.0212} & 0.4673 \scriptsize{$\pm$0.0019} & \textbf{1.0413 \scriptsize{$\pm$0.0021}} & 0.3783 \scriptsize{$\pm$0.1001}\tabularnewline
    ECLB~\cite{eclb} & 0.9213 \scriptsize{$\pm$0.0172} & 0.7812 \scriptsize{$\pm$0.0711} & 0.8918 \scriptsize{$\pm$0.0203} & 0.8329 \scriptsize{$\pm$0.0901} & 0.4913 \scriptsize{$\pm$0.0010} & 0.5239 \scriptsize{$\pm$0.0024} & 1.1028 \scriptsize{$\pm$0.0009} & 0.5912 \scriptsize{$\pm$0.0341}\tabularnewline
    PCFT~\cite{pcft} & 0.8913 \scriptsize{$\pm$0.0108} & 0.8029 \scriptsize{$\pm$0.0692} & \textbf{0.9018 \scriptsize{$\pm$0.0165}} & 0.8495 \scriptsize{$\pm$0.0312} & 0.5982 \scriptsize{$\pm$0.0094} & 0.5671 \scriptsize{$\pm$0.0020} & 1.0913 \scriptsize{$\pm$0.0014} & 0.6824 \scriptsize{$\pm$0.0473}\tabularnewline
    AWGG~\cite{awgg} & 0.9194 \scriptsize{$\pm$0.0081} & 0.8237 \scriptsize{$\pm$0.0124} & 0.9012 \scriptsize{$\pm$0.0081} & 0.8607 \scriptsize{$\pm$0.0135} & \textbf{0.6284 \scriptsize{$\pm$0.0010}} & 0.5564 \scriptsize{$\pm$0.0007} & 1.2130 \scriptsize{$\pm$0.0001} & 0.7209 \scriptsize{$\pm$0.0027}\tabularnewline
    TopoMAD~\cite{topomad} & 0.9229 \scriptsize{$\pm$0.0028} & 0.8562 \scriptsize{$\pm$0.0038} & 0.8927 \scriptsize{$\pm$0.0015} & 0.8741 \scriptsize{$\pm$0.0101} & 0.6098 \scriptsize{$\pm$0.0023} & 0.5330 \scriptsize{$\pm$0.0030} & 1.2892 \scriptsize{$\pm$0.0007} & 0.7313 \scriptsize{$\pm$0.0016}\tabularnewline
    \textbf{DeepFT} & \textbf{0.9422 \scriptsize{$\pm$0.00783}} & \textbf{0.8635 \scriptsize{$\pm$0.0011}} & {0.9001 \scriptsize{$\pm$0.0092}} & \textbf{0.8814 \scriptsize{$\pm$0.0271}} & 0.6193 \scriptsize{$\pm$0.0012} & \textbf{0.5682 \scriptsize{$\pm$0.0072}} &  1.2092 \scriptsize{$\pm$0.0009} & \textbf{0.7556 \scriptsize{$\pm$0.0047}}\tabularnewline
    \bottomrule 
    \end{tabular}} %\vspace{-7pt}
    \label{tab:results}
\end{table*}

\begin{figure*}[!t]
    \centering \setlength\abovecaptionskip{-1pt}
    \setlength\belowcaptionskip{-3pt}
    \subfigure[Energy Consumption per task]{
    \includegraphics[width=.23\textwidth]{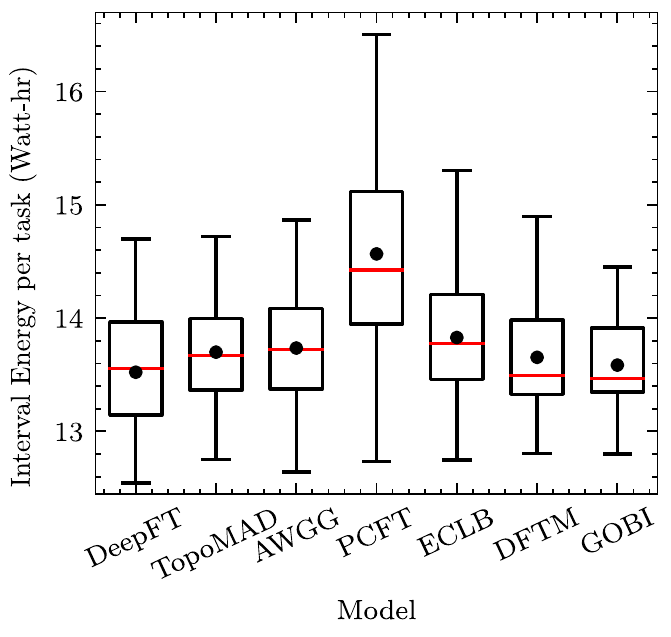}
    \label{fig:energy}
    }
    \subfigure[Number of active containers]{
    \includegraphics[width=.23\textwidth]{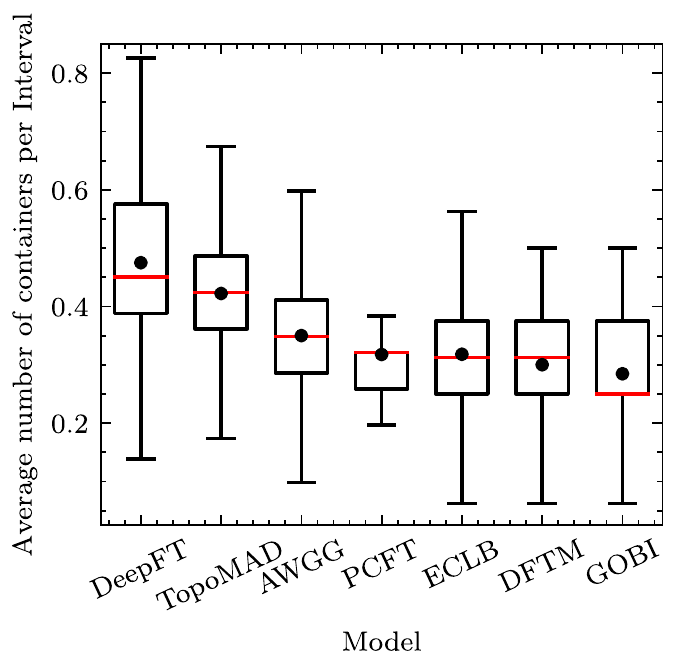}
    \label{fig:num_containers}
    }
    \subfigure[CPU Utilization]{
    \includegraphics[width=.23\textwidth]{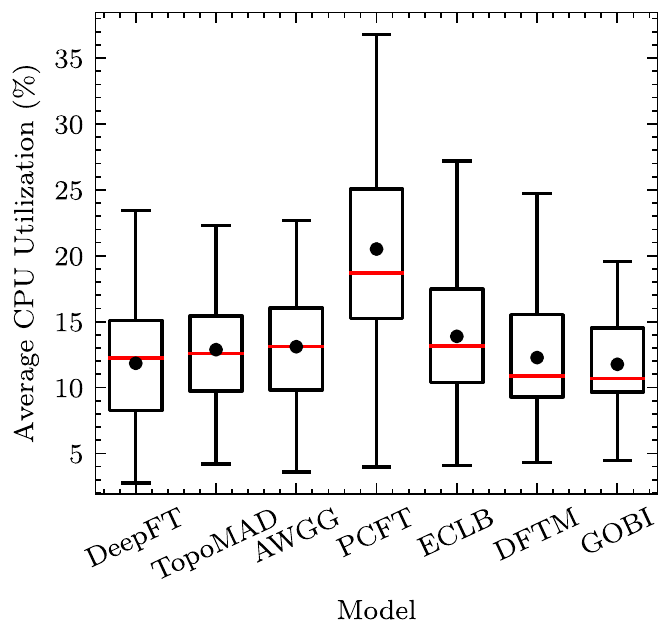}
    \label{fig:cpu}
    }
    \subfigure[RAM Utilization]{
    \raisebox{3pt}{\includegraphics[width=.23\textwidth]{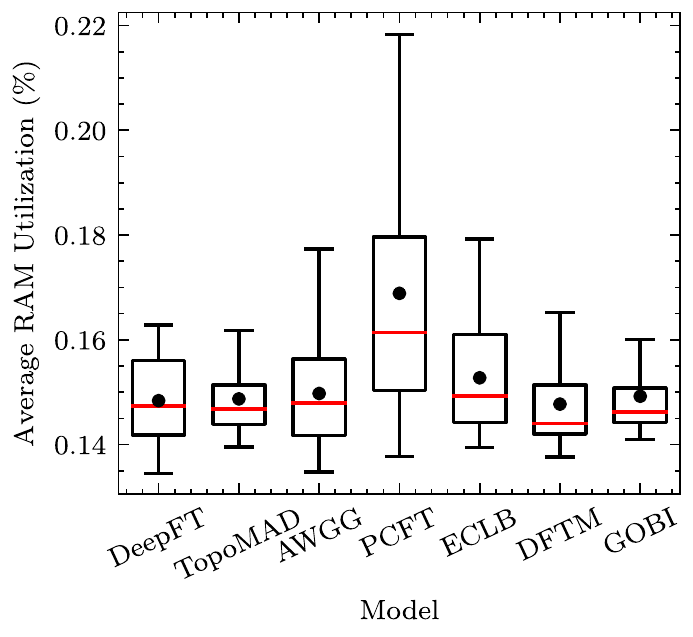}}
    \label{fig:ram}
    }\\ \vspace{-4pt}
    \subfigure[Average Response Time]{
    \includegraphics[width=.23\textwidth]{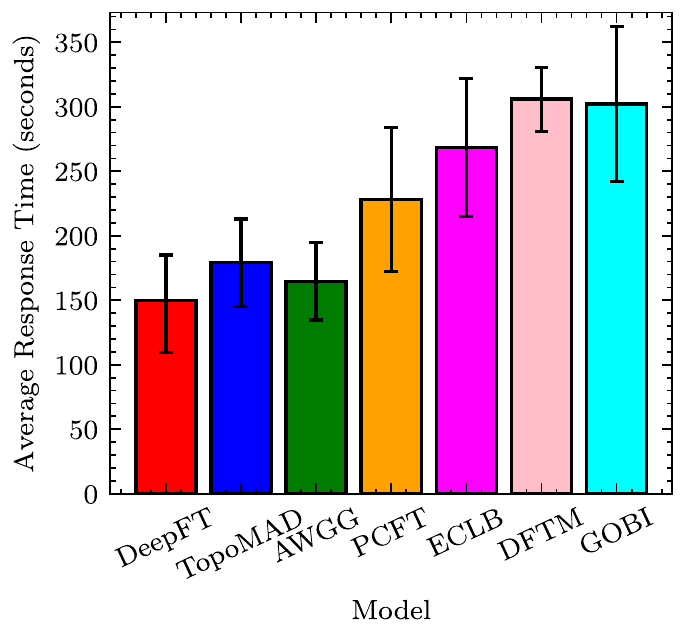}
    \label{fig:rt}
    }
    \subfigure[Average Response Time (per application)]{
    \raisebox{3pt}{\includegraphics[width=.23\textwidth]{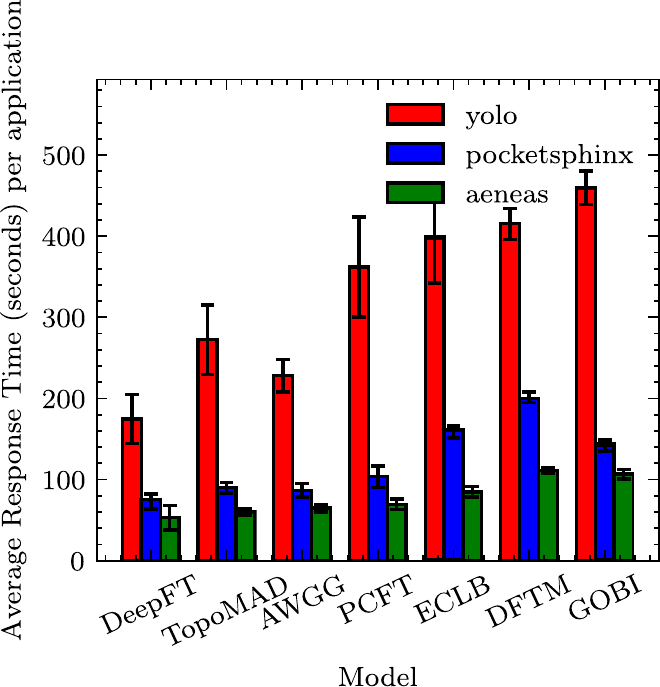}}
    \label{fig:rt_pa}
    }
    \subfigure[Fraction of SLO Violations]{
    \includegraphics[width=.23\textwidth]{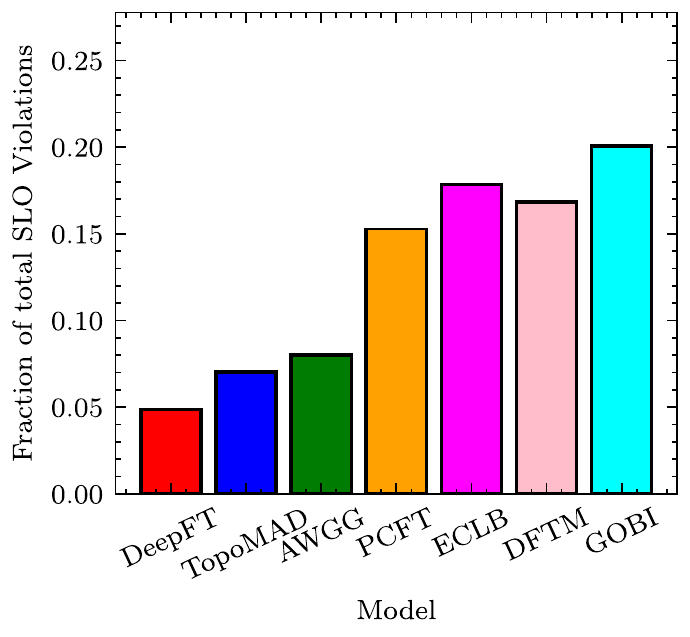}
    \label{fig:sla}
    }
    \subfigure[SLO Violations (per application)]{
    \includegraphics[width=.23\textwidth]{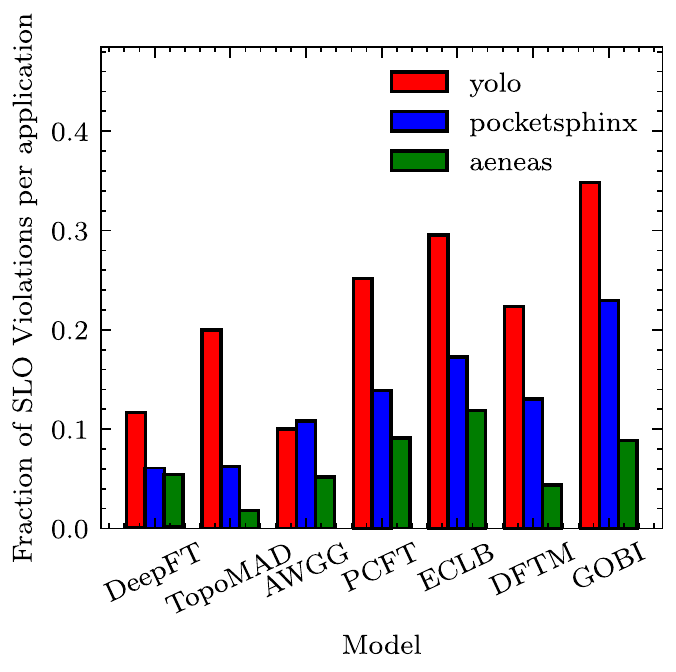}
    \label{fig:sla_pa}
    }\\ \vspace{-4pt}
    \subfigure[Fairness]{
    \raisebox{5pt}{\includegraphics[width=.23\textwidth]{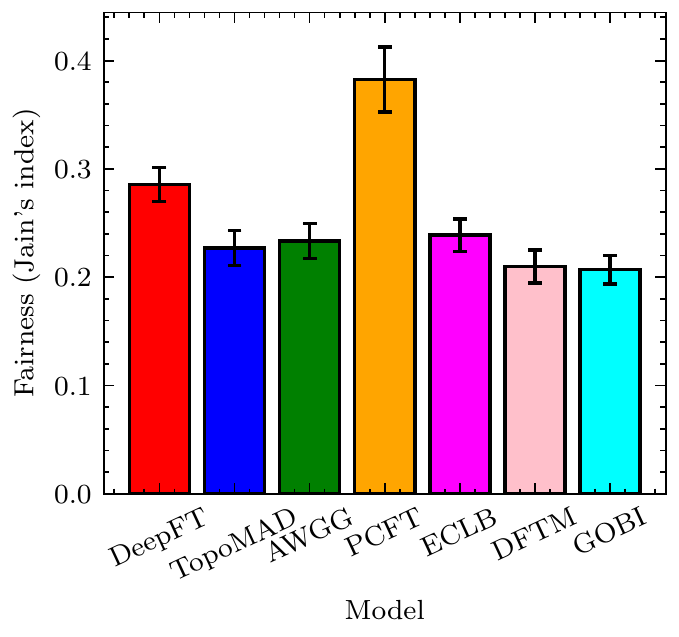}}
    \label{fig:fairness}
    }
    \subfigure[Migration Count]{
    \includegraphics[width=.23\textwidth]{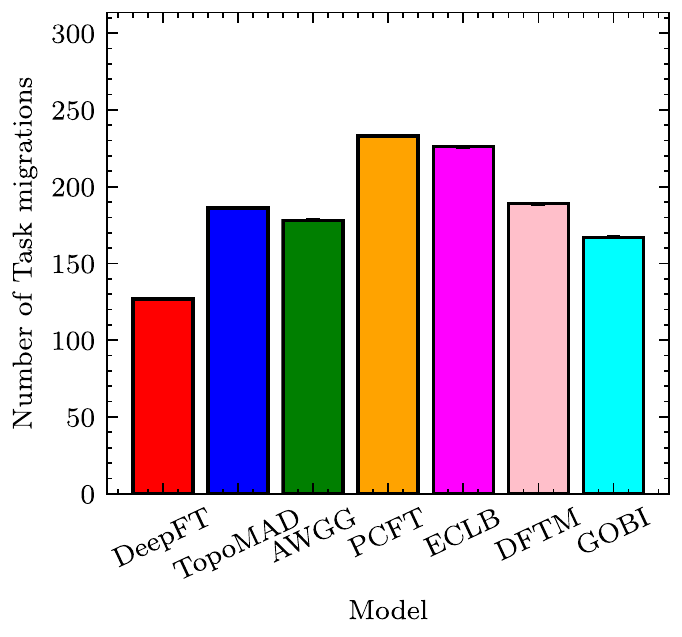}
    \label{fig:migration_count}
    }
    \subfigure[Migration Time vs Interval]{
    \raisebox{12pt}{\includegraphics[width=.23\textwidth]{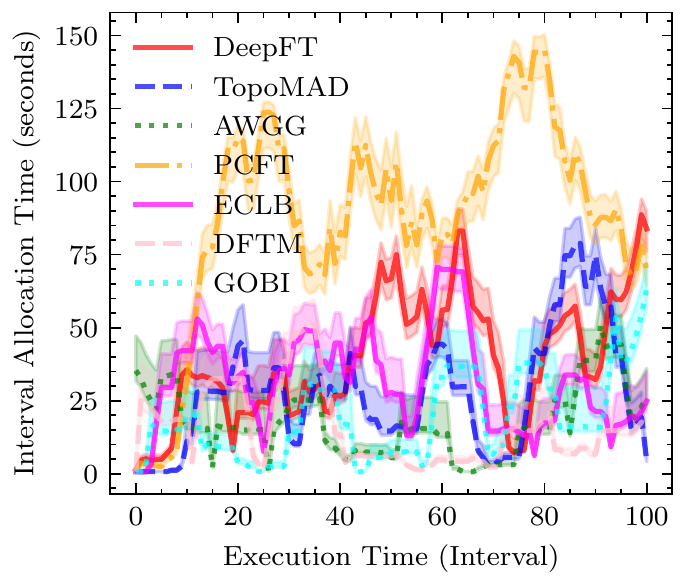}}
    \label{fig:alloc_time}
    }
    \subfigure[Migration Time]{
    \includegraphics[width=.23\textwidth]{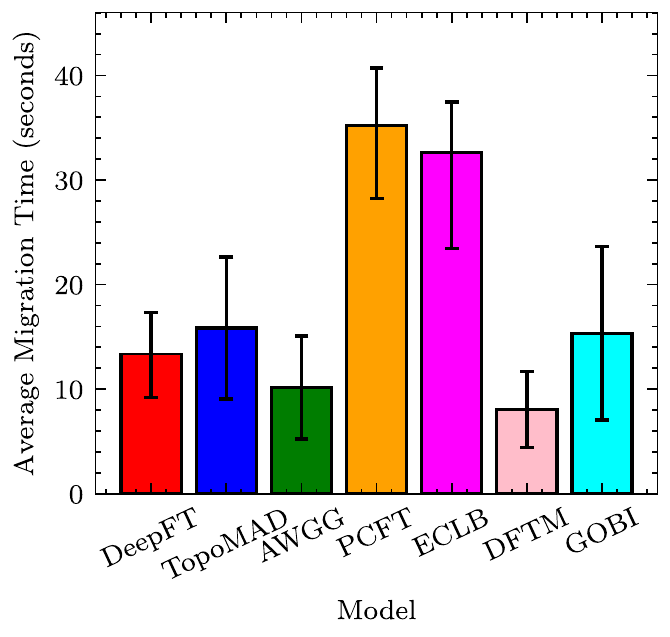}
    \label{fig:migration_time}
    } %\vspace{-2pt}
    \caption{Comparison of QoS parameters of DeepFT against baselines.} 
    \label{fig:results}
\end{figure*}

\subsection{Implementation and Training Details}
To conduct our tests, we extend the COSCO framework that supports Docker based container orchestration in edge environments~\cite{tuli2021cosco}. COSCO is at present the only framework that allows the generation of QoS scores using co-simulated traces. We extend the co-simulation feature with an existing fault model~\cite{nita2014fim}. The resource metrics described in Section~\ref{sec:method} are collected every 10 seconds and the corresponding time-series windows are sent to the broker asynchronously at the start of each scheduling interval. The preemptive migration decisions are extracted as the difference between the input and the output schedules of the DeepFT method. Only those migrations are performed that can be accommodated by the target hosts.

% \blue{In terms of the neural network, we use the $\mathrm{ReLU}$ activation function for all layers. The feed-forward networks are implemented with a single hidden layer with 64 neurons. We use the number of heads of the multi-head attention layers as the size of the sliding window, \textit{i.e.}, same as $k$. The gated graph convolution network is implemented with 4 hidden layers, each of size 64, with $r = 2$ convolutions in equation~\eqref{eq:ggcn}. The GRU layer of size 128 is used to propagate graph updates to the neighboring nodes. The prototype network is implemented as 4 feed-forward layers, each of size 128. All hyperparameters were determined using the \texttt{raytune}\footnote{Ray Tune: \url{https://docs.ray.io/en/latest/tune/index.html}.} Python package. }

To train and fine-tune the surrogate model of the proposed approach, we use the AdamW optimizer~\cite{loshchilov2018decoupled}. We apply a learning rate of $10^{-4}$, number of fault classes $j = 3$ and windows size of $k=5$. We also use a weight regularization parameter of $10^{-4}$. The hyperparameters were obtained using grid-search. Figure~\ref{fig:convergence} shows the convergence plots for the DeepFT's surrogate model. We use the early stopping criterion to train the model. In 21 epochs, the reconstruction and triplet losses converge and we reach detection and classification accuracy of 0.9422 and 0.9592. As this classification does not correspond to any semantic fault class labels, we do not report these results for other models.

\begin{figure*}[!t]
    \centering \setlength\abovecaptionskip{1pt}
    \includegraphics[width=.7\textwidth]{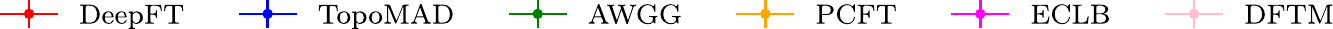}\\
    \subfigure[Energy Consumption]{
    \includegraphics[width=.23\textwidth]{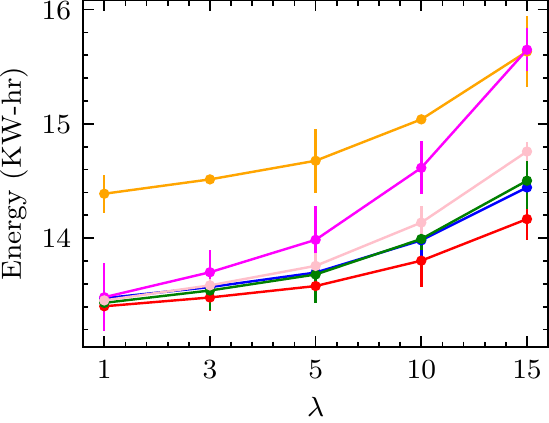}
    \label{fig:s_energy}
    }
    \subfigure[Fraction of SLO Violations]{
    \includegraphics[width=.23\textwidth]{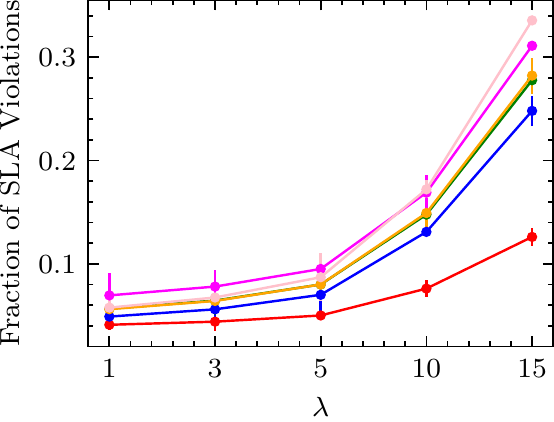}
    \label{fig:s_sla}
    }
    \subfigure[F1 Score]{
    \includegraphics[width=.23\textwidth]{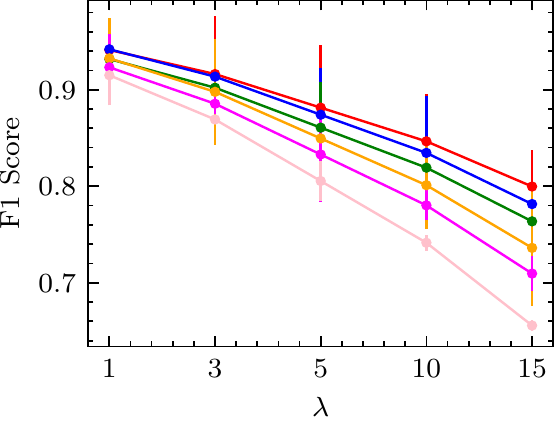}
    \label{fig:s_f1}
    }
    \subfigure[Improvement Ratio]{
    \includegraphics[width=.23\textwidth]{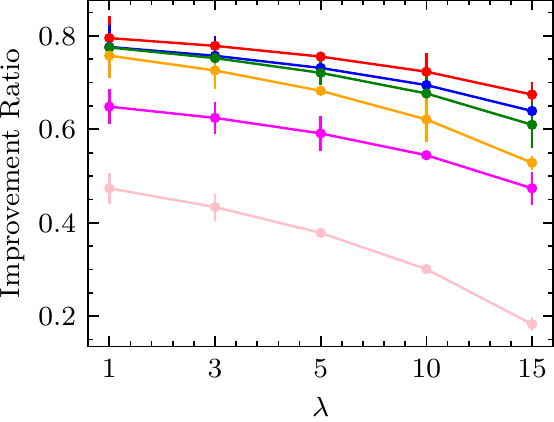}
    \label{fig:s_impr}
    }
    \caption{Sensitivity Analysis for all models with $\lambda$ (parameter of the Poisson distribution).}
    \label{fig:sens}
\end{figure*}

\subsection{Results}

\textbf{Visualization of the DeepFT method.} Figure~\ref{fig:scores} shows how the DeepFT method works starting from a sample scheduling decision. The plot on the left shows the optimization loss $L_O$ with iteration count for the $\mathrm{Adam}$ optimizer. The blips in the optimization curve are due to the warm restarts in cosine annealing. In 20 iterations, the optimization loss drops from 3.03 to 2.43. On the right is a t-SNE plot of the NAP (no-fault) class and the three fault classes. The scheduling decision belongs to the $c_2$ fault class in the beginning, but moves towards the NAP class with iterations. Finally, the converged scheduling decision belongs to the no-fault class shown in red.

\textbf{Comparison with baselines.} Table~\ref{tab:results} shows the detection and diagnosis metrics with the improvement and overhead ratios. Figure~\ref{fig:results} show the QoS metrics of all models for $Poisson(\lambda = 5)$. We also present QoS results for the GOBI scheduler. The fault-detection accuracy of the DeepFT model is the highest (0.9422), improving the best among baseline (0.9229) of the TopoMAD model. Similarly, the DeepFT model also outperforms the baselines in terms of F1 score. The diagnosis scores of the proposed method are close to the state-of-the-art values. \blue{The scores are close to the \textit{supervised} PreGAN method~\cite{tuli2021pregan} without the need for any supervised labels. This demonstrates the efficacy of the DeepFT approach and its capacity to generalize to settings where supervised methods, such as PreGAN, would not be able to perform well due to the lack of ground-truth data.} The improvement ratios of all models except DFTM are higher than 0.5. This  shows that the DFTM model performs slightly worse than the GOBI scheduler in terms of the QoS metrics of response time and energy consumption. However, the SLO violation rates of the DFTM model are lower than GOBI (Fig.~\ref{fig:sla}). This is partly because the DFTM approach performs migrations very aggressively (Fig.~\ref{fig:migration_count}). 

In terms of the improvement ratio, TopoMAD has the highest score of 0.7313 among the baselines. The DeepFT gives a 3.32\% higher score of 0.7556. Even with the periodic fine-tuning of the surrogate model, the overhead ratio of the DeepFT model is 6.20\% lower than the best baseline, TopoMAD. This is due to the relatively time consuming PSO optimization strategy in baselines like TopoMAD and AWGG. DeepFT, on the other hand, uses a directed gradient-based update strategy to optimize the scheduling decision, which has been shown to converge quickly compared to gradient-free methods~\cite{tuli2021cosco}. In terms of the diagnosis results, the AWGG method has the highest HitRate (0.6309), with the DeepFT being very close (0.6193). The NDCG score of the DeepFT model is the highest (0.5682). This is due to the factored fault prediction in the DeepFT model. 

In terms of QoS scores, DeepFT gives the lowest energy consumption per completed task 13.55 KW-hr, with TopoMAD being next with 13.74 KW-hr. This is due to the relatively low average CPU and RAM utilization in DeepFT (Figs.~\ref{fig:cpu}-\ref{fig:ram}) and the high number of active containers (Fig.~\ref{fig:num_containers}). Moreover, DeepFT gives the lowest average response time of 149.2 seconds, with AWGG giving the next best of 164.7 seconds (Fig.~\ref{fig:rt}).  DeepFT also gives a significant reduction of 37.21\% on the SLO violation rates (Fig.~\ref{fig:sla}). This is because  DeepFT avoids unnecessary migrations to prevent avoidable use of network resources, improving overall system reliability (Fig.~\ref{fig:migration_count}) leading to low overall migration times (Fig.~\ref{fig:migration_time}). Figures~\ref{fig:rt_pa} and~\ref{fig:sla_pa} show the average response time and SLO violations for each application. The Yolo application has the highest response time and SLO violation rates due to its computationally heavy requirements. The PCFT method gives the highest fairness index of 0.39, with  DeepFT coming second with 0.28 as shown in Fig.~\ref{fig:fairness}.

\blue{\textbf{Summary.} The results demonstrate that all method for fault-tolerant computing lead to performance improvement, in terms of critical metrics such as SLO violation rates and response times, compared to the case without any fault-tolerance (GOBI). Across all methods, the DeepFT approach gives the best QoS scores, thanks to its high fault prediction accuracy and low overheads.}

\subsection{Sensitivity Analysis}
Figure~\ref{fig:sens} shows the variation of the energy consumption, SLO violations, F1 score and improvement ratio with the $\lambda$ parameter in our Poisson distribution used to model the workloads. We vary $\lambda$ from 1 to 15 ($\lambda = 15$ constantly gives $>90\%$ CPU utilization for all hosts). Under a higher $\lambda$ more tasks are produced, making the fault prediction harder. This is apparent from the drop in the F1 scores, leading to higher SLO violations. Even the energy consumption increases due to the increase in the average CPU utilization of the system. Overall, DeepFT shows the lowest relative drop in F1 scores and improvement ratio as we increase $\lambda$ giving the lowest SLO violations even in workload heavy executions.

\section{Conclusions}
\label{sec:conclusion}

We have presented a deep surrogate model-based fault-tolerance approach (DeepFT) for reliable edge computing. DeepFT can detect and diagnose faulty system conditions without the need for extensive labeled data. Instead, DeepFT utilizes a deep surrogate model that analyzes the scheduling decision and state information to predict a reconstruction of the next state window and fault prototype.  Using reconstruction and few-shot based triplet loss, the surrogate model is trained offline to predict fault scores based on the past data trends. The model represents a self-supervised training model, where it applies co-simulations to generate true state windows. Generating its own ground-truth labels (by utilizing a co-simulator) allows the model to be fine-tuned on-the-fly to adapt it in volatile settings. 

These advances allow DeepFT to have high fault detection and diagnosis scores that facilitate efficient fault-aware scheduling for optimal QoS. Specifically, DeepFT achieves an improvement of 2\% and 3\% for fault-detection accuracy and improvement ratio. DeepFT achieves this with 6.2\% lower overhead than the best baseline TopoMAD. DeepFT is also able to support 1.38\%, 9.41\% and 37.21\% lower energy consumption, response times and SLO violations, respectively compared to the state-of-the-art models. 

Moreover, the parameter size of the surrogate model scales by only 3\% and 1\% with the growth of active tasks and hosts in the system. This makes DeepFT an ideal choice for reliable edge computing with time-critical applications.

As a future work, we plan to explore whether the DeepFT approach can be used for  efficient resource and application management of SLO-based serverless (FaaS) computing at the edge. 
%We now present some future directions for this work. 
Also, considering (in some cases) the availability of limited supervised data, the model may be able to benefit from semi-supervised learning strategies for generalizability. 
%DeepFT can also be extended for serverless provisioning systems that manage edge resources. %Finally, the current model assumes a master-slave design and we plan to explore extensions that allow us to deploy DeepFT in federated or serverless platforms with streaming tasks~\cite{casale2020radon}.

\section*{Software Availability}
The code is available under BSD-3 License at \url{https://github.com/imperial-qore/DeepFT}.

\section*{Acknowledgments}
Shreshth Tuli is supported by the President’s Ph.D. Scholarship at the Imperial College London. 

\bibliographystyle{IEEEtran}

\bibliography{references}

\begin{thebibliography}{10}
\providecommand{\url}[1]{#1}
\csname url@samestyle\endcsname
\providecommand{\newblock}{\relax}
\providecommand{\bibinfo}[2]{#2}
\providecommand{\BIBentrySTDinterwordspacing}{\spaceskip=0pt\relax}
\providecommand{\BIBentryALTinterwordstretchfactor}{4}
\providecommand{\BIBentryALTinterwordspacing}{\spaceskip=\fontdimen2\font plus
\BIBentryALTinterwordstretchfactor\fontdimen3\font minus
  \fontdimen4\font\relax}
\providecommand{\BIBforeignlanguage}[2]{{%
\expandafter\ifx\csname l@#1\endcsname\relax
\typeout{** WARNING: IEEEtran.bst: No hyphenation pattern has been}%
\typeout{** loaded for the language `#1'. Using the pattern for}%
\typeout{** the default language instead.}%
\else
\language=\csname l@#1\endcsname
\fi
#2}}
\providecommand{\BIBdecl}{\relax}
\BIBdecl

\bibitem{narayanan2020key}
A.~Narayanan, A.~S. De~Sena, D.~Gutierrez-Rojas \emph{et~al.}, ``{Key advances
  in pervasive edge computing for industrial Internet of Things in 5G and
  beyond},'' \emph{IEEE Access}, vol.~8, pp. 206\,734--206\,754, 2020.

\bibitem{vasilakos2020towards}
X.~Vasilakos, W.~Featherstone, N.~Uniyal \emph{et~al.}, ``{Towards Zero
  Downtime Edge Application Mobility for Ultra-Low Latency 5G Streaming},'' in
  \emph{2020 IEEE Cloud Summit}.\hskip 1em plus 0.5em minus 0.4em\relax IEEE,
  2020, pp. 25--32.

\bibitem{tuli2022dragon}
S.~Tuli, G.~Casale, and N.~R. Jennings, ``Dragon: Decentralized fault tolerance
  in edge federations,'' \emph{IEEE Transactions on Network and Service
  Management}, 2022.

\bibitem{ray2020proactive}
B.~Ray, A.~Saha, S.~Khatua \emph{et~al.}, ``Proactive fault-tolerance technique
  to enhance reliability of cloud service in cloud federation environment,''
  \emph{IEEE Transactions on Cloud Computing}, 2020.

\bibitem{samanta2021fault}
A.~Samanta, F.~Esposito, and T.~G. Nguyen, ``{Fault-tolerant mechanism for
  edge-based IoT networks with demand uncertainty},'' \emph{IEEE Internet of
  Things}, 2021.

\bibitem{cmodlb}
S.~Negi, M.~M.~S. Rauthan, K.~S. Vaisla \emph{et~al.}, ``{CMODLB: an efficient
  load balancing approach in cloud computing environment},'' \emph{The Journal
  of Supercomputing}, pp. 1--53, 2021.

\bibitem{dftm}
V.~Sivagami and K.~Easwarakumar, ``{An improved dynamic fault tolerant
  management algorithm during VM migration in cloud data center},''
  \emph{Future Generation Computer Systems}, vol.~98, pp. 35--43, 2019.

\bibitem{bagchi2019dependability}
S.~Bagchi, M.-B. Siddiqui, P.~Wood \emph{et~al.}, ``Dependability in edge
  computing,'' \emph{Communications of the ACM}, vol.~63, no.~1, 2019.

\bibitem{goudarzi2020application}
M.~Goudarzi, H.~Wu, M.~Palaniswami \emph{et~al.}, ``{An application placement
  technique for concurrent IoT applications in edge and fog computing
  environments},'' \emph{IEEE Transactions on Mobile Computing}, 2020.

\bibitem{kumar2015fault}
S.~Kumar, D.~S. Rana, and S.~C. Dimri, ``Fault tolerance and load balancing
  algorithm in cloud computing: A survey,'' \emph{International Journal of
  Advanced Research in Computer and Communication Engineering}, vol.~4, no.~7,
  pp. 92--96, 2015.

\bibitem{ledmi2018fault}
A.~Ledmi, H.~Bendjenna, and S.~M. Hemam, ``Fault tolerance in distributed
  systems: A survey,'' in \emph{2018 3rd International Conference on Pattern
  Analysis and Intelligent Systems (PAIS)}.\hskip 1em plus 0.5em minus
  0.4em\relax IEEE, 2018.

\bibitem{beloglazov2012optimal}
A.~Beloglazov and R.~Buyya, ``Optimal online deterministic algorithms and
  adaptive heuristics for energy and performance efficient dynamic
  consolidation of virtual machines in cloud data centers,'' \emph{Concurrency
  and Computation: Practice and Experience}, vol.~24, no.~13, 2012.

\bibitem{eclb}
A.~Sharif, M.~Nickray, and A.~Shahidinejad, ``Fault-tolerant with load
  balancing scheduling in a fog-based iot application,'' \emph{IET
  Communications}, vol.~14, no.~16, pp. 2646--2657, 2020.

\bibitem{tuli2021pregan}
S.~Tuli, G.~Casale, and N.~R. Jennings, ``{PreGAN: Preemptive Migration
  Prediction Network for Proactive Fault-Tolerant Edge Computing},'' in
  \emph{IEEE Conf. on Computer Communications (INFOCOM)}.\hskip 1em plus 0.5em
  minus 0.4em\relax IEEE, 2022.

\bibitem{awgg}
X.~Hu, Y.~Li, L.~Jia \emph{et~al.}, ``{A novel two-stage unsupervised fault
  recognition framework combining feature extraction and fuzzy clustering for
  collaborative AIoT},'' \emph{IEEE Trans. on Industrial Informatics}, 2021.

\bibitem{pcft}
J.~Liu, S.~Wang, A.~Zhou \emph{et~al.}, ``Using proactive fault-tolerance
  approach to enhance cloud service reliability,'' \emph{IEEE Transactions on
  Cloud Computing}, vol.~6, no.~4, pp. 1191--1202, 2016.

\bibitem{topomad}
Z.~He, P.~Chen, X.~Li \emph{et~al.}, ``A spatiotemporal deep learning approach
  for unsupervised anomaly detection in cloud systems,'' \emph{IEEE
  Transactions on Neural Networks and Learning Systems}, 2020.

\bibitem{pecht2019machine}
M.~G. Pecht and M.~Kang, ``Machine learning: Anomaly detection,'' 2019.

\bibitem{goodfellow2016deep}
I.~Goodfellow, Y.~Bengio, and A.~Courville, \emph{Deep learning}.\hskip 1em
  plus 0.5em minus 0.4em\relax MIT press, 2016.

\bibitem{santos2020use}
C.~H.~d. Santos, J.~A. de~Queiroz, F.~Leal \emph{et~al.}, ``Use of simulation
  in the industry 4.0 context: Creation of a digital twin to optimise decision
  making on non-automated process,'' \emph{Journal of Simulation}, pp. 1--14,
  2020.

\bibitem{tuli2022simtune}
S.~Tuli, G.~Casale, and N.~R. Jennings, ``Simtune: bridging the simulator
  reality gap for resource management in edge-cloud computing,''
  \emph{Scientific Reports}, vol.~12, no.~1, pp. 1--12, 2022.

\bibitem{tuli2021cosco}
S.~Tuli, S.~R. Poojara, S.~N. Srirama \emph{et~al.}, ``{COSCO: Container
  Orchestration Using Co-Simulation and Gradient Based Optimization for Fog
  Computing Environments},'' \emph{IEEE Transactions on Parallel and
  Distributed Systems}, vol.~33, no.~1, pp. 101--116, 2022.

\bibitem{kochenderfer2019algorithms}
M.~J. Kochenderfer and T.~A. Wheeler, \emph{Algorithms for optimization}.\hskip
  1em plus 0.5em minus 0.4em\relax Mit Press, 2019.

\bibitem{wang2020generalizing}
Y.~Wang, Q.~Yao, J.~T. Kwok \emph{et~al.}, ``Generalizing from a few examples:
  A survey on few-shot learning,'' \emph{ACM Computing Surveys (CSUR)},
  vol.~53, no.~3, pp. 1--34, 2020.

\bibitem{luo2019improving}
L.~Luo, S.~Meng, X.~Qiu \emph{et~al.}, ``Improving failure tolerance in
  large-scale cloud computing systems,'' \emph{IEEE Transactions on
  Reliability}, 2019.

\bibitem{won2021performance}
H.~Won and Y.~Kim, ``Performance analysis of machine learning based fault
  detection for cloud infrastructure,'' in \emph{2021 International Conference
  on Information Networking (ICOIN)}.\hskip 1em plus 0.5em minus 0.4em\relax
  IEEE, 2021, pp. 877--880.

\bibitem{li2018comparison}
C.~Li, M.~Cerrada, D.~Cabrera \emph{et~al.}, ``A comparison of fuzzy clustering
  algorithms for bearing fault diagnosis,'' \emph{Journal of Intelligent \&
  Fuzzy Systems}, vol.~34, no.~6, pp. 3565--3580, 2018.

\bibitem{audibert2020usad}
J.~Audibert, P.~Michiardi, F.~Guyard \emph{et~al.}, ``{USAD: UnSupervised
  Anomaly Detection on Multivariate Time Series},'' in \emph{Proc. of the 26th
  ACM SIGKDD Intl. Conf. on Knowledge Discovery \& Data Mining}, 2020, pp.
  3395--3404.

\bibitem{tanenbaum1997operating}
A.~S. Tanenbaum and A.~S. Woodhull, \emph{Operating systems: design and
  implementation}.\hskip 1em plus 0.5em minus 0.4em\relax Prentice Hall
  Englewood Cliffs, 1997, vol.~68.

\bibitem{javed2018cefiot}
A.~Javed, K.~Heljanko, A.~Buda \emph{et~al.}, ``{CEFIoT: A fault-tolerant iot
  architecture for edge and cloud},'' in \emph{2018 IEEE 4th world forum on
  internet of things (WF-IoT)}.\hskip 1em plus 0.5em minus 0.4em\relax IEEE,
  2018, pp. 813--818.

\bibitem{liu2018partial}
G.~Liu, F.~A. Reda, K.~J. Shih \emph{et~al.}, ``Image inpainting for irregular
  holes using partial convolutions,'' in \emph{The European Conference on
  Computer Vision (ECCV)}, 2018.

\bibitem{snell2017prototypical}
J.~Snell, K.~Swersky, and R.~Zemel, ``Prototypical networks for few-shot
  learning,'' in \emph{Proceedings of the 31st International Conference on
  Neural Information Processing Systems}, 2017, pp. 4080--4090.

\bibitem{aima}
S.~Russell and P.~Norvig, \emph{Artificial Intelligence: A Modern Approach},
  3rd~ed.\hskip 1em plus 0.5em minus 0.4em\relax USA: Prentice Hall Press,
  2009.

\bibitem{ioffe2015batch}
S.~Ioffe and C.~Szegedy, ``Batch normalization: Accelerating deep network
  training by reducing internal covariate shift,'' in \emph{International
  conference on machine learning}.\hskip 1em plus 0.5em minus 0.4em\relax PMLR,
  2015, pp. 448--456.

\bibitem{vaswani2017attention}
A.~Vaswani, N.~Shazeer, N.~Parmar \emph{et~al.}, ``Attention is all you need,''
  in \emph{Proceedings of the 31st International Conference on Neural
  Information Processing Systems}, 2017, pp. 6000--6010.

\bibitem{ruiz2020gated}
L.~Ruiz, F.~Gama, and A.~Ribeiro, ``Gated graph recurrent neural networks,''
  \emph{IEEE Transactions on Signal Processing}, vol.~68, pp. 6303--6318, 2020.

\bibitem{bahdanau2015neural}
D.~Bahdanau, K.~H. Cho, and Y.~Bengio, ``Neural machine translation by jointly
  learning to align and translate,'' in \emph{3rd International Conference on
  Learning Representations, ICLR 2015}, 2015.

\bibitem{siffer2017anomaly}
A.~Siffer, P.-A. Fouque, A.~Termier \emph{et~al.}, ``Anomaly detection in
  streams with extreme value theory,'' in \emph{ACM SIGKDD International
  Conference on Knowledge Discovery and Data Mining}, 2017.

\bibitem{medina2020selfsupervised}
C.~Medina, A.~Devos, and M.~Grossglauser, ``{Self-Supervised Prototypical
  Transfer Learning for Few-Shot Classification},'' \emph{International
  Conference on Machine Learning (ICML) - Workshop on Automated Machine
  Learning}, 2020.

\bibitem{kingma2014adam}
D.~P. Kingma and J.~Ba, ``Adam: A method for stochastic optimization,''
  \emph{International Conference on Learning Representations}, 2015.

\bibitem{loshchilov2016sgdr}
I.~Loshchilov and F.~Hutter, ``{SGDR: Stochastic gradient descent with warm
  restarts},'' \emph{arXiv preprint arXiv:1608.03983}, 2016.

\bibitem{paszke2019pytorch}
A.~Paszke, S.~Gross, F.~Massa \emph{et~al.}, ``Pytorch: An imperative style,
  high-performance deep learning library,'' \emph{Advances in Neural
  Information Processing Systems}, vol.~32, pp. 8026--8037, 2019.

\bibitem{spec}
\BIBentryALTinterwordspacing
{Standard Performance Evaluation Corporation}. {SPEC Power Consumption Models}.
  [Online]. Available: \url{https://www.spec.org/cloud\_iaas2018/results/}
\BIBentrySTDinterwordspacing

\bibitem{mcchesney2019defog}
J.~McChesney, N.~Wang, A.~Tanwer \emph{et~al.}, ``{DeFog: fog computing
  benchmarks},'' in \emph{The 4th ACM/IEEE Symp. on Edge Computing}, 2019, pp.
  47--58.

\bibitem{mao2016dynamic}
Y.~Mao, J.~Zhang, and K.~B. Letaief, ``Dynamic computation offloading for
  mobile-edge computing with energy harvesting devices,'' \emph{IEEE Journal on
  Selected Areas in Communications}, 2016.

\bibitem{basu2019learn}
D.~Basu, X.~Wang, Y.~Hong \emph{et~al.}, ``Learn-as-you-go with megh: Efficient
  live migration of virtual machines,'' \emph{IEEE Transactions on Parallel and
  Distributed Systems}, vol.~30, no.~8, pp. 1786--1801, 2019.

\bibitem{boettiger2015introduction}
C.~Boettiger, ``An introduction to docker for reproducible research,''
  \emph{ACM SIGOPS Operating Systems Review}, vol.~49, no.~1, pp. 71--79, 2015.

\bibitem{morabito2017evaluating}
R.~Morabito, I.~Farris, A.~Iera \emph{et~al.}, ``Evaluating performance of
  containerized iot services for clustered devices at the network edge,''
  \emph{IEEE Internet of Things Journal}, vol.~4, no.~4, pp. 1019--1030, 2017.

\bibitem{ade}
\BIBentryALTinterwordspacing
{Open Mainframe Project}. {Anomaly Detection Engine (ADE) for Linux Logs}.
  [Online]. Available:
  \url{https://www.openmainframeproject.org/projects/anomaly-detection-engine-for-linux-logs-ade}
\BIBentrySTDinterwordspacing

\bibitem{omnianomaly}
Y.~Su, Y.~Zhao, C.~Niu \emph{et~al.}, ``Robust anomaly detection for
  multivariate time series through stochastic recurrent neural network,'' in
  \emph{Proc. of the 25th ACM SIGKDD Intl. Conf. on Knowledge Discovery \& Data
  Mining}, 2019, pp. 2828--2837.

\bibitem{jarvelin2002cumulated}
K.~J{\"a}rvelin and J.~Kek{\"a}l{\"a}inen, ``Cumulated gain-based evaluation of
  ir techniques,'' \emph{ACM Trans. on Information Systems (TOIS)}, vol.~20,
  no.~4, pp. 422--446, 2002.

\bibitem{nita2014fim}
M.-C. Nita, F.~Pop, M.~Mocanu \emph{et~al.}, ``Fim-sim: fault injection module
  for cloudsim based on statistical distributions,'' \emph{Journal of
  telecommunications and information technology}, no.~4, pp. 14--23, 2014.

\bibitem{loshchilov2018decoupled}
I.~Loshchilov and F.~Hutter, ``Decoupled weight decay regularization,'' in
  \emph{International Conference on Learning Representations}, 2018.

\end{thebibliography}

\end{document}